\documentclass[11pt,preprint]{pasj00}
\SetRunningHead{Murata \& Matsubara}{Impacts of smoothing kernels.}
\Received{2006/09/04}
\Accepted{2006/12/06}
\Published{2007/??/??}

\draft

\begin{document}
\title{Effects of Smoothing Functions in Cosmological Counts-in-Cells Analysis}
\author{Yoshitaka \textsc{Murata}${}^{1}$ and Takahiko
  \textsc{Matsubara}${}^{1,2}$}
\affil{${}^{1}$ Department of Physics and Astrophysics, Nagoya University,
  Chikusa, Nagoya 464-8602, Japan}
\affil{${}^{2}$ Institute for Advanced Research, Nagoya University,
  Chikusa, Nagoya, 464-8602, Japan}

\email{murata@a.phys.nagoya-u.ac.jp, taka@nagoya-u.ac.jp}

\KeyWords{cosmology: cosmological parameters --- cosmology: large-scale
  structure of universe --- cosmology: theory --- methods: analytical}
\maketitle

\begin{abstract}
    A method of counts-in-cells analysis of galaxy distribution is
    investigated with arbitrary smoothing functions in obtaining the
    galaxy counts. We explore the possiblity of optimizing the
    smoothing function, considering a series of $m$-weight
    Epanechnikov kernels. The popular top-hat and Gaussian smoothing
    functions are two special cases in this series. In this paper, we
    mainly consider the second moments of counts-in-cells as a first
    step. We analytically derive the covariance matrix among different
    smoothing scales of cells, taking into account possible overlaps
    between cells. We find that the Epanechnikov kernel of $m=1$ is
    better than top-hat and Gaussian smoothing functions in estimating
    cosmological parameters. As an example, we estimate expected
    parameter bounds which comes only from the analysis of second
    moments of galaxy distributions in a survey which is similar to
    the Sloan Digital Sky Survey.
\end{abstract}

\section{Introduction}

The large-scale structure of the universe is one of the most powerful
probes of the universe. The structure of the galaxy distribution is a
consequence of how our universe began, how it has been evolved with
time, and what the universe is made of. Galaxy distributions are
inherently statistical, and methods of quantifying them are not
uniquely given. The counts-in-cells (CIC) analysis, which we consider
in this paper, is one of the most simple methods among them. After
quantifying the observed galaxy distributions by some statistical
quantity, the cosmological parameters are estimated by that quantity.

According to the observations of the temperature anisotropy of the
cosmic microwave background radiation (CMB), such as the Wilkinson
microwave anisotropy probe (WMAP) \citep{wmap,wmap3} and
magnitude-redshift relation of type Ia SNe (SNIa)
\citep{1aSNe1,1aSNe2}, the cosmological constant is necessary to
account for the behavior of the universe. However, in order to
understand the evolution of structures, it is also important to
explore structures in different epochs. Galaxy distributions probe
more recent universe than observations such as CMB and SNIa. Recent
galaxy redshift surveys, such as the AAT Two-Degree Field Galaxy
Redshift Survey (2dFGRS) and the Sloan Digital Sky Survey (SDSS)
reveal the large-scale structure of the mostly present universe
\citep{2df2,sdss}. By combining the data of the galaxy distributions
and other observations like WMAP, many cosmological parameters are
accurately estimated \citep{2df-para,sdss-para,LySDSS}.

One of recent trends of the parameter estimation is to use a number of
independent observations in order to improve constraints of the
parameters as accurate as possible. Other one is to constrain the
cosmological parameters from independent observations in order to make
cross-checks of a cosmological model. The latter one is important
because the assumed cosmological model is not guaranteed to be
correct. For example, the dark energy is introduced just as a
parameter, not knowing the deep nature. On the other hand, there are
many efforts to explain an acceleration of the universe without
introducing dark energy. People are trying to explain the acceleration
of the universe by, for example, a back reaction due to super-horizon
density fluctuations, modifications of gravity, and so on
\citep{superhorizon-I,mond1,superhorizon-review,mond2,mond3,superhorizon-no}.

In estimating cosmological parameters from independent observations,
some of the parameters cannot be determined with good accuracy because
of possible weak signals or degeneracies between parameters. For
example, when we consider the cosmological parameter estimation from
the CMB, the free streaming dumping due to neutrinos is less
significant than that in the present large-scale structure. When we
consider the cosmological parameter estimation from the large-scale
structure, the silk dumping due to baryons and the free streaming due
to neutrinos show similar dumping in the power spectrum on small
scales. Each observation excels others at estimating some parameters
but is not good at estimating other parameters.

In this paper, we consider parameter estimations using only a CIC
analysis of the large-scale structure. In the CIC analysis, one needs
a smoothing function, which is a function of distance from a center of
each cell. Usually, the top-hat function and the Gaussian function are
used. The top-hat function is often used for counting galaxies in each
cells. The Gaussian smoothing function is used when we need to smooth
out small-scale clustering and to obtain large-scale properties of
clustering. For example, \citet{non-tophat} used a Gaussian-like
smoothing function for that purpose. Although top-hat and Gaussian
functions are simple, they are not necessarily optimal for estimating
cosmological parameters. In this study, we investigate effects of the
choice of smoothing functions. We introduce a series of smoothing
functions: the $m$-weight Epanechnikov kernels. The top-hat function
and the Gaussian function are two special cases in this series. Since
this function has intermediate properties between these two functions,
we expect that it is useful to search for better function in analyzing
the large-scale structure. Since the $m$-weight Epanechnikov kernel
has a finite support, its Fourier counterpart of the kernel,
$\tilde{W}_m(k)$, oscillates as a function of $k$. This property leads
us to expect a possibility that we can find a specific smoothing
function which is sensitive to oscillating features in the power
spectrum, such as the baryon acoustic oscillations (BAO).

This paper is organized as follows. In \S~\ref{sec:cic-epa}, we
briefly summarize the CIC analysis and the $m$-weight Epanechnikov
kernels. In \S~\ref{sec:fisher}, our methods using the covariance
matrix in analytic form with the Fisher information matrix are
explained. In \S~\ref{sec:results}, we present numerical results of
our study. Conclusions and discussion are given in
\S~\ref{sec:discussion}.

\section{A Counts-in-Cells Analysis and Smoothing Functions}
\label{sec:cic-epa}

\subsection{A Counts-in-Cells Analysis}

We consider a CIC analysis of the galaxy distribution on large scales.
The CIC analysis is one of the methods to analyze the large-scale
structure. First of all, a number of cells are randomly distributed in
a survey volume and the number of galaxies are counted in each cell.
Next, the moments, or semi-invariants, are calculated from the numbers
of galaxies in those cells. The CIC analysis has an advantage that it
is simpler to analyze than more complex statistics such as the
two-point correlation function and the power spectrum, etc. Behaviors
of the statistical errors are also studied well \citep{lap}.

The CIC analysis was developed by \citet{peebles} and applied to the
IRAS and Stromo-APM redshift survey \citep{iras,stromo-apm,cic}. In
these papers, they measured a count probability distribution function,
$P_N(R)$, which is a probability that a cell contains $N$ galaxies in
a spherical cell. They obtained the second moment of the distribution
function from likelihood fitting of the distribution function to an
analytic log-normal function. Recently, a skewed log-normal function
is used as an improved analytic function \citep{SLND,skewedL,SLNM}. In
this study, we simply consider the second moments of the galaxy
counts, which directly obtained by averaging the square of galaxy
counts \citep{peebles}.

The CIC analysis is related to the smoothed density field with a
smoothing function $W(r;R)$:
\begin{equation}
  \delta_R(\boldsymbol{r})
  =
  \frac{3}{4\pi R^3}\int d^3r' \,
  W\left(|\boldsymbol{r} - \boldsymbol{r}'|; R \right)
  \delta(\boldsymbol{r}'),
\label{eq:10}
\end{equation}
where $\delta(\boldsymbol{r})$ is a 3-dimensional density contrast and
$R$ is a smoothing scale. In this paper, we use a normalization
\begin{equation}
  \int d^3r \, W\left(|\boldsymbol{r}|; R\right) = \frac{4\pi R^3}{3}.
\label{eq:20}
\end{equation}
This normalization is adopted in such a way that top-hat smoothing
function has a normalization, $W_{\rm TH}\left(r;R\right) = 1\ (r<R)$,
$ 0\ (r>R)$. When the function is not a top-hat one, this
normalization is just a convention. The second moment of the density
contrast is given by \citep{statistics}
\begin{eqnarray}
  \sigma^2(R) \equiv \left\langle {\delta_R}^2 \right\rangle
  &=&
  \left(\frac{3}{4\pi R^3}\right)^2 \int d^3r_1 d^3r_2 \,
  \xi\left(|\boldsymbol{r}_1-\boldsymbol{r}_2|\right)
  W\left(|\boldsymbol{r}_1|;R\right)W\left(|\boldsymbol{r}_2|;R\right)
  \nonumber\\
  &=&
  \left(\frac{3}{4\pi R^3}\right)^2
  \int\frac{d^3k}{(2\pi)^3} P(k)|\tilde{W}(kR)|^2
\label{eq:variance},
\end{eqnarray}
where $\xi(r)$ is the two-point correlation function, $P(k)$ is the
power spectrum, and
\begin{equation}
  \tilde{W}(kR) \equiv
  \int d^3r \, e^{-i\boldsymbol{k}\cdot\boldsymbol{r}}
  W\left(|\boldsymbol{r}|; R\right)
\label{eq:30}
\end{equation}
is a Fourier counterpart of the smoothing function, which is a real
function for a spherical smoothing function.  Similarly, the
higher-order moments, $\left\langle {\delta_R}^3 \right\rangle$,
$\left\langle {\delta_R}^4 \right\rangle$, $\ldots$ are given by
higher-order correlation functions and polyspectra.

The density contrast $\delta(\boldsymbol{r})$ is a smooth function of
space, while the count of galaxy is a discrete number. We consider the
case that galaxies are counted in each cell with a weight function
$W(|\boldsymbol{r}|; R)$, which is a same function we introduced above
as a smoothing function. Each galaxies are counted with a weight
according to the distance from centers of cells. The count of galaxies
$N$ in a cell is defined by
\begin{equation}
  N = \sum_{g: {\rm galaxies}}
  W\left(|\boldsymbol{r}_g - \boldsymbol{r}_{\rm c}|; R \right),
\label{eq:40}
\end{equation}
where $\boldsymbol{r}_g$ is a position of each galaxies, and
$\boldsymbol{r}_c$ is a center of the cell. For a top-hat smoothing
function, the above quantity exactly corresponds to a number of
galaxies in a sphere of radius $R$, and $N$ is a non-negative
integer. In general smoothing functions, $N$ is not necessarily an
integer, despite its notation. Hereafter we call $N$ a weighted count.

In a case of the top-hat smoothing function, relations between moments
of the smoothed density contrast and moments of the galaxy counts are
given in \citet{peebles}. We need to generalize such relations to the
cases of general smoothing function. We derive those relations in
Appendix~\ref{Ap:moments-counts}. The second moment of the density
contrast is related to the galaxy counts by equation~(\ref{eq:sigma}),
\begin{equation}
  \sigma^2(R) =
  \frac{\langle N^2\rangle}{\langle N\rangle^2} -
  \frac{\overline{W}}{\langle N\rangle} - 1,
\label{eq:50}
\end{equation}
where $\langle N\rangle$ is the weighted count averaged over all
cells, $\langle N^2\rangle$ is the second moment of the weighted
count, and
\begin{equation}
  \overline{W} \equiv \frac{3}{4\pi R^3}
  \int d^3r \, \left\{ W\left(|\boldsymbol{r}|;R\right) \right\}^2.
\label{eq:60}
\end{equation}
The right hand side of equation~(\ref{eq:50}) are given only by
observable quantities. The second term is a shot noise term. Relations
between higher-order moments of the density contrast and that of
weighted counts are given in the Appendix~\ref{Ap:moments-counts},
although we focus only on the second moments below.

\subsection{The $m$-weight Epanechnikov kernel}

The choice of the smoothing function $W(r;R)$ remains arbitrary in the
analysis explained above. The top-hat function
\begin{equation}
  W_{\rm TH}(r;R) = \Theta(R-r),
\label{eq:70}
\end{equation}
where
\begin{equation}
  \Theta(x) = \left\{
  \begin{array}{ll}
    0, & (x<0), \\
    1, & (x>0),
  \end{array}
  \right.
\label{eq:80}
\end{equation}
is a Heaviside step function, is usually adopted in the CIC
analysis. Another choice is a Gaussian function,
\begin{equation}
  W_{\rm G}(r;R) =
  \frac{2}{3\sqrt{2\pi}}\exp\left(-\frac{r^2}{2R^2}\right).
\label{eq:90}
\end{equation}
In this study, we introduce a series of smoothing functions which has
an intermediate properties of the two: the $m$-weight Epanechnikov
kernel. This series of functions are defined by
\begin{equation}
  W_m(r;R) =
  \frac{(2m+3)!!}{3\cdot 2^m\, m!}
  \left(1-\frac{r^2}{R^2}\right)^m\Theta(R-r).
\label{eq:100}
\end{equation}
In this smoothing function with positive weight $m$, the weight is
large near the center of cells and small near the edge of cells. In
the case $m=0$, this kernel is the top-hat function of
equation~(\ref{eq:70}). The original Epanechnikov kernel
\citep{epaorg} corresponds to $m=1$, and the $m$-weight kernel of $m
\ne 0,1$ is a generalized kernel of the original one. The Fourier
counterparts of the Epanechnikov kernels are given by
\begin{equation}
  \tilde{W}_m(kR) =
  \frac{4\pi R^3}{3}
  \frac{(2m+3)!!j_{m+1}(kR)}{(kR)^{m+1}}.
\label{eq:110}
\end{equation}
Since the $m$-weight Epanechnikov kernel has a finite support, its
Fourier counterpart $\tilde{W}_m(k)$ oscillates as a function of $k$.
One can expect a possibility that there is a specific smoothing
function which is sensitive to a certain physical feature in the power
spectrum, such as the baryon acoustic oscillations (BAO).

In the $m$-weight Epanechnikov kernels with a large weight $m$, the
weight near the edge of the cells, $r \sim R$ is very small. Therefore
the characteristic radius of the kernel is not given by the radius of
the support $R$. Instead, it is useful to define an effective radius
$R_{\rm eff}$ of the $m$-weight Epanechnikov kernel as a
characteristic radius:
\begin{equation}
  R_{\rm eff} =
  \left[\frac{3}{4\pi R^3}\int d^3r\,
  W_m\left(|\boldsymbol{r}|;R\right)\,r^2\right]^{1/2}
  = \left(\frac{3}{2m+5}\right)^{1/2}R.
\label{eq:reff}
\end{equation}
When we take the limit of $m\rightarrow\infty$ fixing the effective
radius $R_{\rm eff}$, the $m$-weight Epanechnikov kernel is shown to
be reduced to the Gaussian function $W_{\rm G}(r;R_{\rm eff})$
\citep{epa-fim}. Therefore, the series of $m$-weight Epanechnikov
kernels contain both the top-hat function ($m=0$) and the Gaussian
function ($m = \infty$) as special cases. Although taking a simple
limit of $m \rightarrow \infty$ results in divergent expressions in
this paper because of our normalization convention, the derived
observable quantities are still regular in this limit.

\section{The Fisher Matrix Analysis and the Covariance Matrix}
\label{sec:fisher}

\subsection{The Fisher matrix analysis}

The main purpose of this study is to find an appropriate smoothing
function. Since the degrees of freedom in choosing the smoothing
function are infinitely large, we constrain ourselves in searching an
optimal weight $m$ of the $m$-weight Epanechnikov kernels. To this
end, we use the method of Fisher information analysis. This analysis
is a method to theoretically estimate expected parameter bounds that
can be achieved with an unbiased estimator \citep{CRinequality}. This
analysis is often used in order to estimate how well one can constrain
cosmological parameters in a given observation
\citep[etc.]{fim-one-sigma,fim-mcmc,future}.

The Fisher information matrix $\boldsymbol{F}$ is defined by
\begin{equation}
  \boldsymbol{F}_{ij}(\boldsymbol{\theta}) = -
  \left\langle
  \frac{\partial^2\ln \mathcal{L}(\boldsymbol{\theta}|\boldsymbol{x})}
    {\partial\theta_i\partial\theta_j}
  \right\rangle,
\label{eq:120}
\end{equation}
where $\mathcal{L}(\boldsymbol{\theta}|\boldsymbol{x})$ is a
likelihood function of model parameters $\boldsymbol{\theta}$ when a
set of data $\boldsymbol{x}$ is given.  The average is taken over
assumed distribution of data, which is specified by a fiducial model
with a set of parameters $\boldsymbol{\theta}$.  When the likelihood
function has the maximum at a set of fiducial model parameters, the
first derivatives of the likelihood should vanishes. The second
derivatives specify how the likelihood function is peaked at the
maximum point. Thus, the Fisher matrix is a useful quantity to
estimate how one can constrain model parameters in a likelihood
analysis.

In fact, when only one parameter $\theta_i$ is estimated, fixing other
parameters, there is a Cram\'er-Rao inequality
\citep{CRinequality,fim1,epa-fim},
\begin{equation}
  \left\langle (\Delta\theta_i)^2 \right\rangle
  \geq \left(F_{ii}\right)^{-1}
\label{eq:bound}
\end{equation}
where $\Delta\theta_i \equiv \theta_i - \langle \theta_i \rangle$ is a
deviation of that parameter, and the right hand side of
equation~(\ref{eq:bound}) is a square of a standard deviation. In the
limit of a very large data set, the inequality becomes an
equality. Therefore, the diagonal elements of a Fisher matrix offer
criteria of how powerful a given analysis is. When multiple parameters
are simultaneously estimated, there is a generalized correspondence,
\begin{equation}
  \left\langle
    \Delta\theta_i \Delta\theta_j
  \right\rangle
  \sim
  \left(\boldsymbol{F}^{-1}\right)_{ij},
\end{equation}
in the limit of a very large data set. Therefore, expected parameter
bounds by a given data set can be estimated by calculating a Fisher
matrix.

According to the Bayes' theorem, the likelihood function in
equation~(\ref{eq:120}) is given by
\begin{equation}
  \mathcal{L}(\boldsymbol{\theta}|\boldsymbol{x})
  = \frac{\mathcal{P}(\boldsymbol{x}|\boldsymbol{\theta})
  \mathcal{P}(\boldsymbol{\theta})}{\mathcal{P}(\boldsymbol{x})},
\label{eq:130}
\end{equation}
where $\mathcal{P}(\boldsymbol{x}|\boldsymbol{\theta})$ is the
conditional probability distribution function of the data
$\boldsymbol{x}$ for a given set of parameters $\theta$,
$\mathcal{P}(\boldsymbol{\theta})$ is the prior distribution function
of parameters. The prior distribution function of data
$\mathcal{P}(\boldsymbol{x})$ plays a role of just a normalization in
the Bayes' theorem.

We assume in the below that there is not any prior knowledge of the
parameters. In that case, the prior distribution function
$\mathcal{P}(\boldsymbol{\theta})$ is constant, and
$\mathcal{L}(\boldsymbol{\theta}|\boldsymbol{x}) \propto
\mathcal{P}(\boldsymbol{x}|\boldsymbol{\theta})$ as functions of
parameters $\boldsymbol{\theta}$. In equation~(\ref{eq:120}),
$\mathcal{L}(\boldsymbol{\theta}|\boldsymbol{x})$ can be replaced by
$\mathcal{P}(\boldsymbol{x}|\boldsymbol{\theta})$. When the
distribution of data $\boldsymbol{x}$ is given by a multivariate
Gaussian function,
\begin{equation}
  \mathcal{P}(\boldsymbol{x}|\boldsymbol{\theta}) =
  \frac{1}{\sqrt{(2\pi)^N \det \boldsymbol{C}}}
  \exp\left[
    -\frac{1}{2}(\boldsymbol{x} - \langle\boldsymbol{x}\rangle)^{\rm t}
    \boldsymbol{C}^{-1}(\boldsymbol{x}-\langle\boldsymbol{x}\rangle)\right],
\label{eq:lf}
\end{equation}
where $N$ is the dimension of the data vector $\boldsymbol{x}$, i.e.,
the number of data. The $N\times N$ matrix $\boldsymbol{C}$ is a
covariance matrix given by
\begin{equation}
  \boldsymbol{C}(\boldsymbol{\theta}) =
  \left\langle
    (\boldsymbol{x} - \langle\boldsymbol{x} \rangle)
    (\boldsymbol{x} - \langle\boldsymbol{x} \rangle)^{\rm t}
  \right\rangle,
\label{eq:cova}
\end{equation}
where the average is taken over data $\boldsymbol{x}$, which
distribution is given by a set of model parameters
$\boldsymbol{\theta}$. Thus, the covariance matrix explicitly depends
on model parameters $\boldsymbol{\theta}$. The Fisher matrix for a
multivariate Gaussian distribution of equation~(\ref{eq:lf}) is given
by \citep{fim2}
\begin{equation}
  F_{ij}(\boldsymbol{\theta}) =
  \frac{1}{2}\left[\boldsymbol{C}^{-1}
    \frac{\partial\boldsymbol{C}}{\partial\theta_i}\boldsymbol{C}^{-1}
    \frac{\partial\boldsymbol{C}}{\partial\theta_j}\right]
  + \frac{\partial\langle\boldsymbol{x}\rangle^t}
  {\partial\theta_i}\boldsymbol{C}^{-1}
  \frac{\partial\langle\boldsymbol{x}\rangle}{\partial\theta_j}.
\label{eq:140}
\end{equation}

In reality, the distribution of data is generally not a multivariate
Gaussian. Even in that case, Fisher matrix of equation~(\ref{eq:140})
is still useful to estimate the expected parameter bounds. In fact,
\citet{fim-mcmc} compares the Fisher information matrix analysis and
the Markov Chain Monte Carlo (MCMC) method by using a simulation.
According to their results, the Fisher matrix analysis and MCMC method
agrees well, when the likelihood function is not much different from a
(multivariate) Gaussian function. When the likelihood function is
sufficiently peaked at the maximum, the Fisher matrix analysis gives a
good prediction for expected parameter bounds.

\subsection{The covariance matrix of second moments}

The Fisher matrix is calculated by equation~(\ref{eq:140}), once a
covariance matrix $\boldsymbol{C}$ is obtained as a function of model
parameters. We use a second moment of the CIC as a set of data
$\boldsymbol{x}$. For a fixed smoothing function, the second moments
$\sigma^2(R)$ is a function of a smoothing radius, $R$. We assume that
we have a set of data which consists of second moments with various
$R$, and the model parameters are estimated by a maximum likelihood
analysis. Therefore, the data vector is given by
\begin{equation}
  \boldsymbol{x} = 
  \left[
    \sigma^2(R_1), \sigma^2(R_2), \ldots, \sigma^2(R_N)
  \right]^{\rm t},
\label{eq:150}
\end{equation}
where $R_1 < R_2 < \cdots < R_N$ are a given set of smoothing radii.
The values of the smoothing radii are chosen so that the density
fluctuations are modeled by linear theory and we treat the problem as
analytic as possible. Thus the minimum value of the smoothing radius
$R_1$ cannot be arbitrarily small.  On the other hand, the smoothing
radius cannot be arbitrarily large because the radius cannot be larger
than a survey size.

The number of data, which is the same as the number of smoothing
radii, can be arbitrary chosen. The larger the number of data is, the
tighter the parameters are constrained. However, second moments with
slightly different smoothing radii do not have quite independent
information. This property is taken into account in our analysis as we
will see below.

The matrix elements of the covariance matrix we need is therefore
given by
\begin{equation}
  C_{ij} = 
  \left\langle
    \left[
      \sigma^2(R_i) -
      \left\langle \sigma^2(R_i) \right\rangle
    \right]
    \left[
      \sigma^2(R_j) -
      \left\langle \sigma^2(R_j) \right\rangle
    \right]
  \right\rangle.
\label{eq:160}
\end{equation}
The calculation of this covariance matrix is not trivial. We develop a
method to analytically calculate this matrix under certain
approximations, which is technically a main elaboration of this work.
The details are described in Appendix~\ref{Ap:covariance-matrix}. The
statistical distribution of the second moments depends on the
individual properties of galaxy samples, such as the density of
galaxies, the size of the survey volume, etc. These properties are
taken into account in the calculation of the covariance matrix. In the
calculation, we allow the volume of a cell can be overlapped to that
of another cell. Therefore, the number of cells placed in the survey
volume is not limited. We adopt an approximation that a survey volume
has a spherical shape with a radius $L$. The boundary effects of the
survey volume are neglected. These approximations are useful for an
analytical tractability of the calculation
(Appendix~\ref{Ap:covariance-matrix}).

\section{Numerical Results}
\label{sec:results}

\subsection{Assumed galaxy surveys and cosmological parameters}
\label{sec:fiducial}

The first sample we consider is the one which is similar to the main
galaxy sample of SDSS. We refer this sample as MG. In this sample, the
number density of galaxies is set to be $\overline{n} = 3.3\times
10^{-3}\,(h^{-1}{\rm Mpc})^{-3}$. The survey size is set by a radius
$L = 378\,h^{-1}{\rm Mpc}$. The second sample is the one which is
similar to the luminous red galaxy sample of SDSS. We refer this
sample as LRG. This sample is characterized by $\overline{n} = 10^{-4}
\, (h^{-1}{\rm Mpc})^{-3}$ and $L = 723 \, h^{-1} {\rm Mpc}$. The
redshifts of galaxies in both samples are not large and we simply
neglect redshift evolution of galaxy clustering.  We also neglect the
bias for simplicity, although the real samples in the SDSS has a
certain bias. The QSO sample in the SDSS has very sparse density
\citep{epa-fim} and we do not consider a sample of this kind. We do
not mean to make MG and LRG to resemble the actual SDSS
survey. Instead, we intend to contrast two different types of sample.
Since we neglect the redshift evolution of clustering and the bias,
the power spectrum of both samples are identical. The main difference
is the survey volume and the number density, which are decisive
factors in cosmological parameter estimations.

In the Fisher analysis, we need to assume maximum likelihood estimates
of parameters, which is called fiducial model parameters. We adopt the
fiducial parameters as $(\Omega_{\rm m}, f_{\rm B}, \Omega_{\rm \nu},
h, n_{\rm s}, {\sigma_8}^2) = (0.3, 0.15, 0., 0.7, 1.0, 1.0)$, where
$\Omega_{\rm m}$ is the matter density parameter, $f_B$ is a fraction
of the density against the matter density, $\Omega_\nu$ is the
neutrino density parameter, $h$ is the hubble constant normalized by
$100 {\rm km/s/Mpc}$, $n_s$ is the spectral index of the primordial
power spectrum, and ${\sigma_8}^2$ is the standard normalization of the
density fluctuations. The bias factor of both samples, MG and LRG, are
assumed to be unity. In order to obtain the power spectrum of the
present universe, we calculate the transfer function by the CMBfast
code \citep{cmbfast}. Normalization of the power spectrum is
determined by setting ${\sigma_8}^2$. To ensure that the typical scale
of fluctuations should not be in strongly nonlinear regime, we only
use effective smoothing radii of equation~(\ref{eq:reff}) which
satisfies $R_{\rm eff} \ge 10\,h^{-1}{\rm Mpc}$. The largest smoothing
radii should be fairly smaller than the survey size, since we neglect
boundary effects of the survey volume. In our calculation below, the
largest radii are chosen as $R_{\rm eff} = 75\,h^{-1}{\rm Mpc}$ for
the MG and $R_{\rm eff} = 110\,h^{-1}{\rm Mpc}$ for the LRG. Linearly
equal spacings of smoothing radii $R_i$ in equation~(\ref{eq:150}) are
adopted. The number of smoothing radii $N$ are varied in the following
analysis. The number of cells placed in the survey volume $N_{\rm
  cells}$ are given by $(L/R_{\rm eff})^3$.

\subsection{Effects of smoothing function}

\begin{figure}[h]
  \begin{center}
    \FigureFile(140mm,100mm){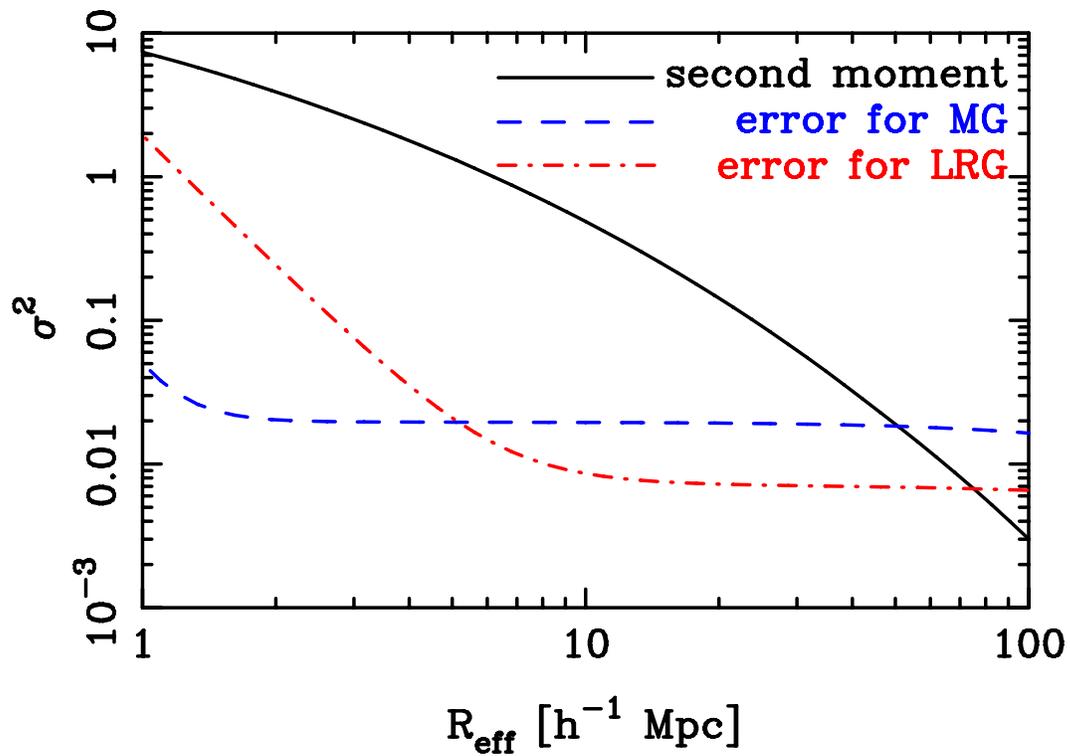}
  \end{center}
  \caption{The second moment with a
    top-hat smoothing function and its errors expected from Fisher
    matrix calculation. {\it Solid line}: a common second moment of
    both samples in the fiducial model; {\it dashed line}: expected
    $1\sigma$ error for the MG; {\it dot-dashed line}: expected
    $1\sigma$ error for the LRG.
    \label{fig:fig1}}
\end{figure}

In figure~\ref{fig:fig1}, we show the second moments $\sigma^2$ and
their statistical errors as functions of $R_{\rm eff}$ for both the MG
and the LRG. These are analytically given by
equations~(\ref{eq:variance}) and (\ref{eq:B21}). In this figure, the
solid line is the second moment of the fiducial model, which is common
to both samples, dashed line is an expected error for the MG and
dot-dashed line is an error for the LRG. Relatively flat errors on
larger scales are dominated by cosmic variance, which are less for the
LRG than the MG because of the larger sampling volume. A little
decrease of the error in MG around $100\,h^{-1}{\rm Mpc}$ is due to
the boundary effect which is not properly taken into account in our
analytic calculation.  On scales where the boundary effect is
significant, errors due to the cosmic variance is already larger than
the signal, and we do not have to worry about the boundary
effect. Increasing errors toward smaller scales are due to the shot
noise, which is less for the MG because the MG is denser than the
LRG. As mentioned above, we only use the large scales, $R_{\rm eff}
\geq 10\, h^{-1}{\rm Mpc}$, where the errors are dominated by cosmic
variance, and thus the LRG has larger signal to noise ratio than that
of the MG.

In figure~\ref{fig:fig1}, there is a little decrease in the cosmic
variance at $\sim100\,h^{-1}{\rm Mpc}$ in the case of the MG sample .
This is caused by our approximation to neglect the boundary effects as
described in Appendix~\ref{Ap:covariance-matrix}. Therefore, when $R$
becomes comparable to $L$, the cosmic variance is underestimated.
However, the signal is already dominated by the cosmic variance in
this regime, and this effect does not affect the results.

The number of cells placed in the survey volume $N_{\rm cell}$ are
given by $(L/R_{\rm eff})^3$, as mentioned in the previous section.
The number affects the results through a factor $1/N_{\rm cell}$
appearing in equation~(\ref{eq:B21}). The first line of
equation~(\ref{eq:B21}) represents the shot noise, the second line the
shot noise from overlapping regions, and the third line the cosmic
variance. In increasing the numbers of cells for both the MG and the
LRG, the shot noises decrease as mentioned by
\citet{non-tophat}. However, the shot noise from overlapping regions
becomes dominant on small scales $R_{\rm eff}\le10$ $h^{-1}$ Mpc. In
addition, the statistical erros are dominated by the cosmic variances
on $R_{\rm eff}\ge10$ $h^{-1}$ Mpc, and the factor appearing in the
third line of equation~(\ref{eq:B21}) is $\lesssim$ 0.01 at most.
Therefore, we can find from equation~(\ref{eq:B21}) that the numbers
of cells are sufficient and that increasing the number of cells does
not affect the following results: the statistical error does not
diminish more than $1\%$ if the number of cells is increased.

On the other hand, in decreasing the number of cells, scales dominated
by the shot noise become larger, and the signal to noise ratio
diminishes. For the MG, decreasing the number of cells to one tenth of
the present number does not affect the signal to noise ratio so much.
However, decreasing the number to one hundredth significantly lowers
the ratio. Therefore, the number of cells is enough for the cosmic
variance to dominate the shot noise on $R_{\rm eff}=10$ $h^{-1}$ Mpc.
The shot noise contributes the error only $\lesssim1\%$ on that scale.
For the LRG, decreasing to one tenth significantly reduces the signal
to noise ratio, since the shot noise is comparable to the cosmic
variance on scales of $R_{\rm eff}\sim10$ $h^{-1}$ Mpc. A hundred
times larger number of cells than the present number is necessary to
suppress the shot noise contribution to the same level as the MG on
$R_{\rm eff}=10$ $h^{-1}$ Mpc. Therefore, increasing the number of
cells enlarges the signal to noise ratio on scales of $R_{\rm
  eff}\sim10$ $h^{-1}$ Mpc and would slightly tighten constraints of
the parameters.

\begin{figure}[h]
  \begin{center}
    \FigureFile(84mm,60mm){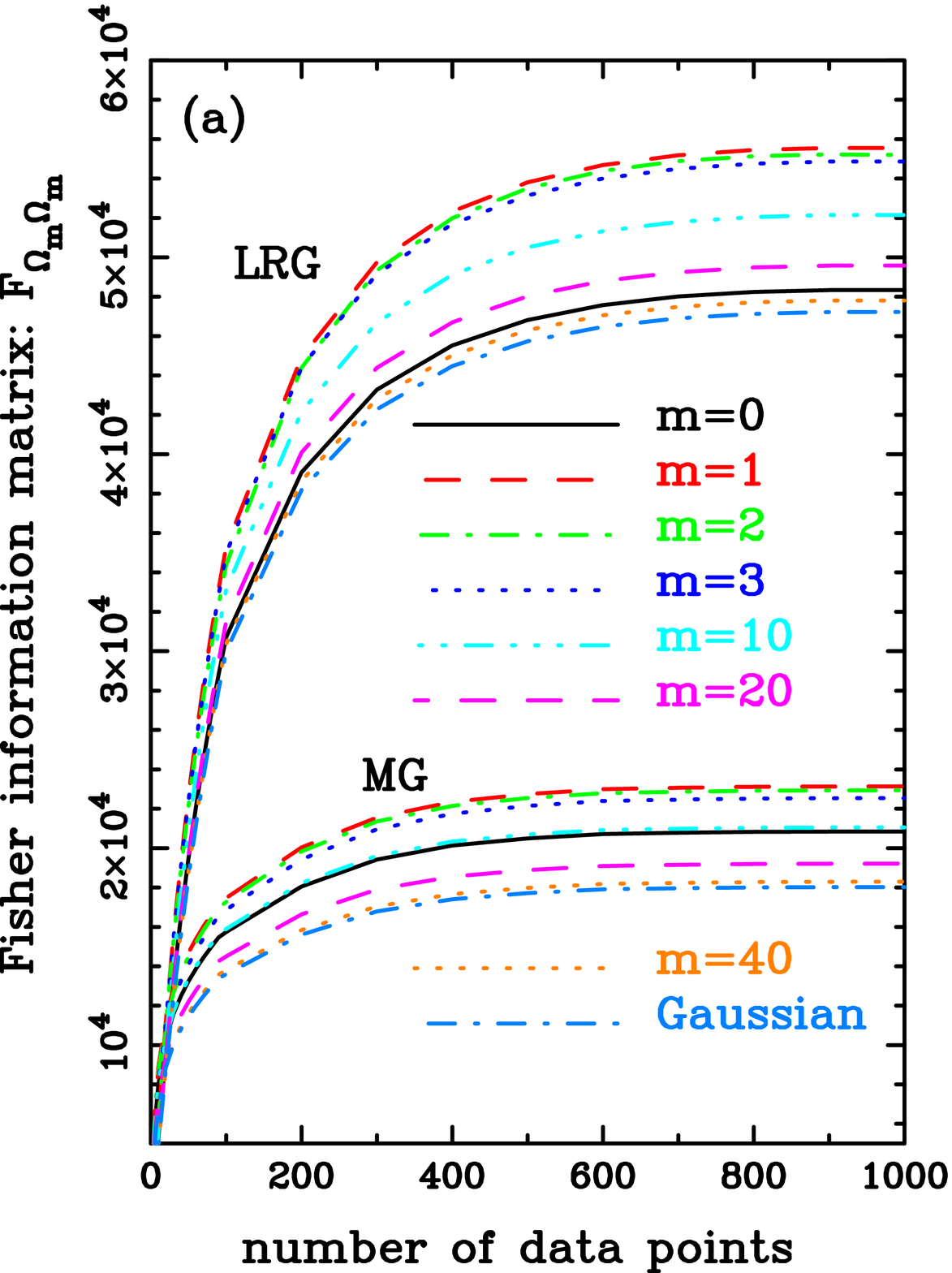}
    \FigureFile(84mm,60mm){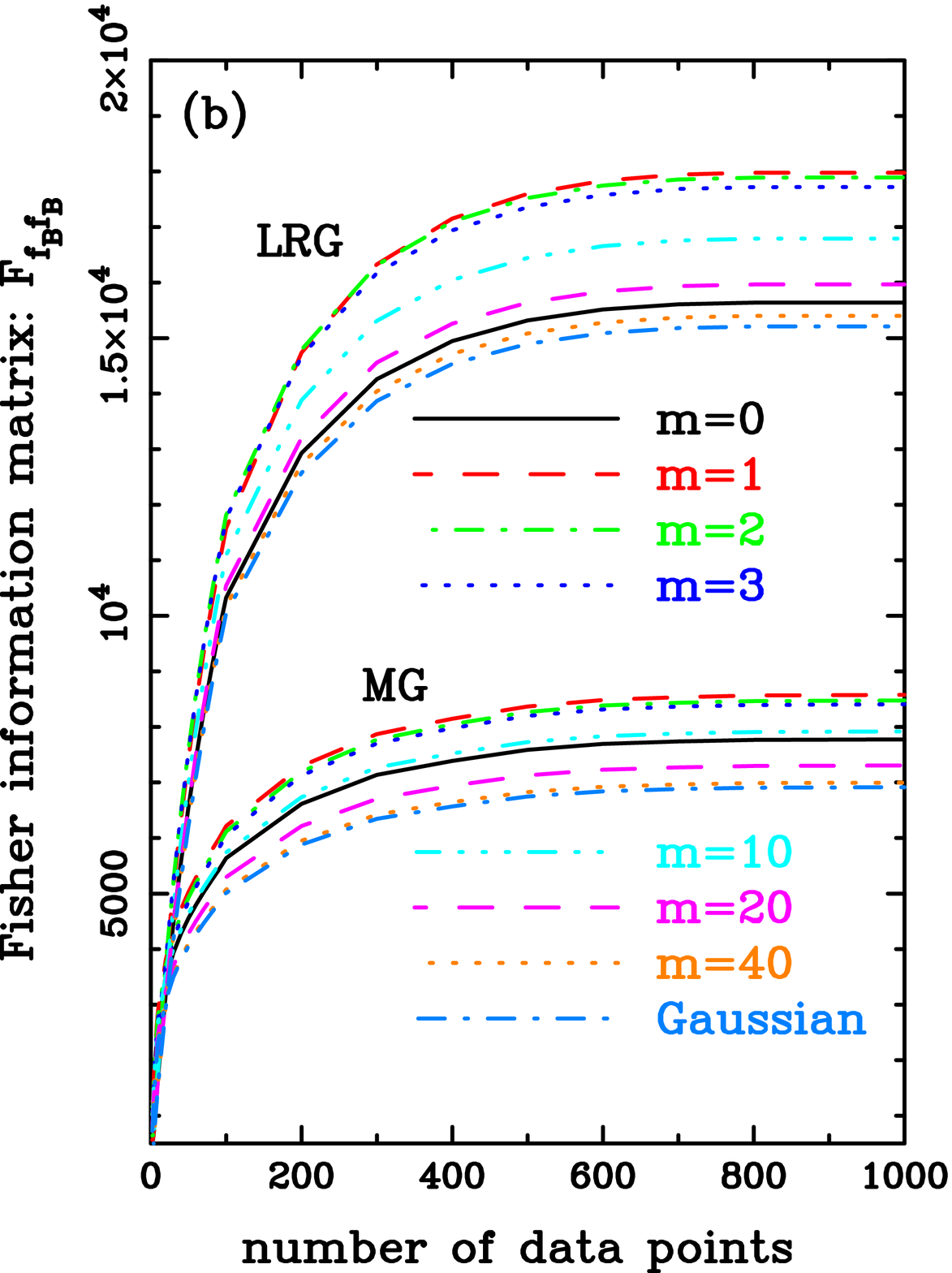}
  \end{center}
  \caption{{\it Panel~(a)}: dependences of the matter density
    parameter element of the Fisher information matrix on the number of
    data points and on the various weights of the $m$-weight
    Epanechnikov kernel for both the MG sample and LRG sample. {\it
      Panel~(b)}: dependences of the baryon fraction parameter. The
    upper eight lines are for the LRG sample and the lower eight lines
    are for the MG sample. For each sample, {\it solid line}: $m=0$;
    upper {\it dashed line}: $m=1$; upper {\it dot-dashed line}: upper
    $m=2$; {\it dotted line}: $m=3$; {\it dot-dot-dot-dashed line}:
    $m=10$; lower {\it dashed line}: $m=20$; lower {\it dotted line}:
    $m=40$; lower {\it dot-dashed line}: Gaussian window function.
    \label{fig:fig2}}
\end{figure}

Figure~\ref{fig:fig2} shows how the matter density parameter and the
baryon fraction are constrained by plotting the diagonal elements of
the corresponding Fisher matrix as functions of the numbers of data,
i.e., the number of smoothing radius $N$. This number can be
arbitrarily large because this number is artificially chosen. For
larger numbers of data, the Fisher diagonal elements are larger and
expected error bounds become smaller. However, the Fisher diagonal
elements cease to increase for sufficiently large numbers of data,
because the second moments of similar smoothing radii do not carry
independent information. Therefore, the saturation values of the Fisher
diagonal elements for the large numbers of data indicate the expected
error bound from the analysis of second moments.

According to this figure, the MG with about 400 data points is enough
to obtain nearly minimum errors for both $\Omega_{\rm m}$ and $f_{\rm
  B}$. That means increasing the numbers of data above 400 is not
efficient. In the LRG, about 600 data points are enough. The LRG can
constrain the cosmological parameters more tightly than the MG sample,
because of the large volume. For both the MG and the LRG, expected
errors for other parameters have similar tendencies.

The saturation values are different for different $m$, which shows
effects of smoothing functions for the parameter estimation. The
Fisher diagonal elements in both parameters are largest when we use
$m=1$ (original) Epanechnikov kernel. Using the top-hat and Gaussian
functions are nearly worst among $m$-weight Epanechnikov kernels.
Improvements of the expected errors are about 14\% for the LRG and
11\% for the MG (in each case). Although the improvement is not so
significant, using the Epanechnikov kernel is definitely better than
using top-hat and Gaussian kernels as a smoothing function. The reason
why the Epanechnikov kernel is better than top-hat and Gaussian kernel
will be discussed in the next section.

There is another suggestion in this Figure. For example, expected
errors for $N=400$ by the top-hat ($m=0$) function are almost the same
as $N=200$ by the $m=1$ Epanechnikov function in every case. In this
case, optimizing the smoothing function has the same effect as
increasing the data points with respect to the parameter estimation.
In practice, increasing data points increases computational costs of
evaluating the likelihood function. Using the optimized smoothing
function has an advantage in this respect.

\begin{figure}[h]
  \begin{center}
    \FigureFile(140mm,100mm){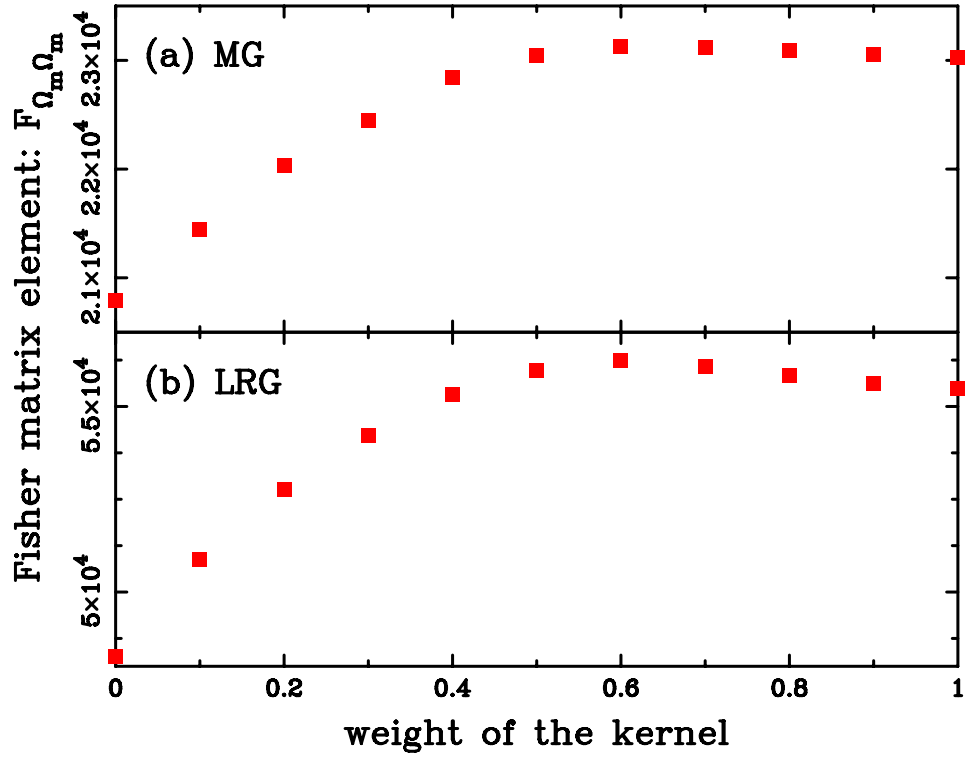}
  \end{center}
  \caption{The matter density element of the Fisher information matrix
    for a non-integer weight kernel. {\it Panel~(a)}: the MG sample;
    {\it Panel~(b)}: the LRG sample. \label{fig:fig6}}
\end{figure}

In Figure~\ref{fig:fig2}, the expected errors for weights $m=0$ and
$m=1$ are quite different in both the MG and the LRG. It is natural to
consider non-integer weights to looking for more optimal function.
Figure~\ref{fig:fig6} shows the dependence of the Fisher diagonal
elements for $\Omega_{\rm m}$ against the weight $m$. 
It follows from this figure that $m = 0.6$
kernel is slightly better than the $m=1$ kernel for both the MG and
the LRG.

\begin{figure}[h]
  \begin{center}
    \FigureFile(140mm,100mm){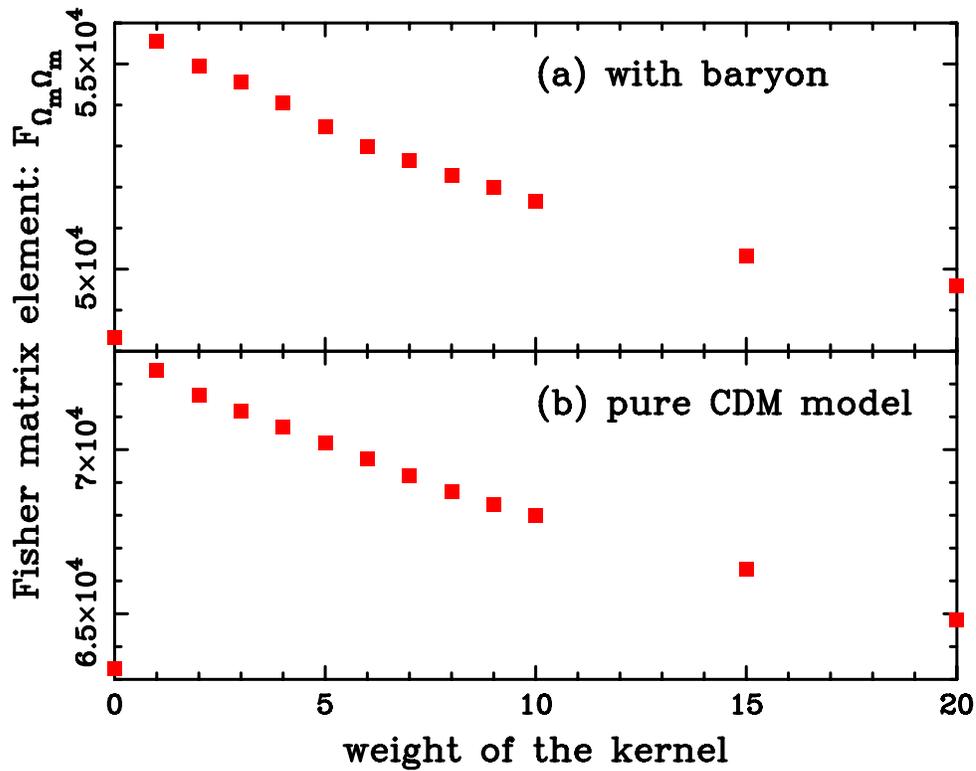}
  \end{center}
  \caption{Dependences of the Fisher matrix element
    $F_{\Omega_m\Omega_m}$ on weight of the kernel. {\it Panel~(a)}:
    the maximum likelihood point is assumed to be the fiducial model;
    {\it panel~(b)}: the maximum likelihood point is assumed to be
    $(\Omega_{\rm m}, f_{\rm B}, \Omega_{\rm \nu}, h, n_{\rm s},
    {\sigma_8}^2) = (0.3, 0., 0., 0.7, 1.0,
    1.0)$.  \label{fig:nobaryon}}
\end{figure}

To test the effect of BAO in the power spectrum, we compare the Fisher
matrix elements with and without the baryonic component in the power
spectrum. Figure~\ref{fig:nobaryon} shows a result. Panel~(a) shows a
dependence of the value of Fisher matrix element
$F_{\Omega_m\Omega_m}$ with the baryonic component. The fiducial model
of the panel~(a) is the same as that explained in
\S~\ref{sec:fiducial}. Panel~(b) shows a same dependence as the
panel~(a) but in the pure CDM model. The fiducial parameters in this
panel are assumed to be $(\Omega_{\rm m}, f_{\rm B}, \Omega_{\rm \nu},
h, n_{\rm s}, {\sigma_8}^2) = (0.3, 0., 0., 0.7, 1.0, 1.0)$. In the
panel~(b), we use the fitting formula of \citet{pureCDM} to calculate
the transfer function in the pure CDM model. If the BAO plays a main
role in the improvement of parameter constraints, it is expected that
the pure CDM model does not show any dependences on weight of the
kernels, which is not found in this study. Therefore, we conclude that
the improvements come from other reasons, such as an overall envelope
of the smoothing function or covariances. In addition, if the
oscillation feature is the main reason, we expect the constraint for
baryon fraction parameter is especially tightened, which is also not
found in this study. Therefore, the fact that the Epanechnikov kernel
of $m=1$ gives the tightest constraints for every parameter also
supports this conclusion.

\subsection{Expected parameter bounds}

\begin{figure}[h]
  \begin{center}
    \FigureFile(140mm,100mm){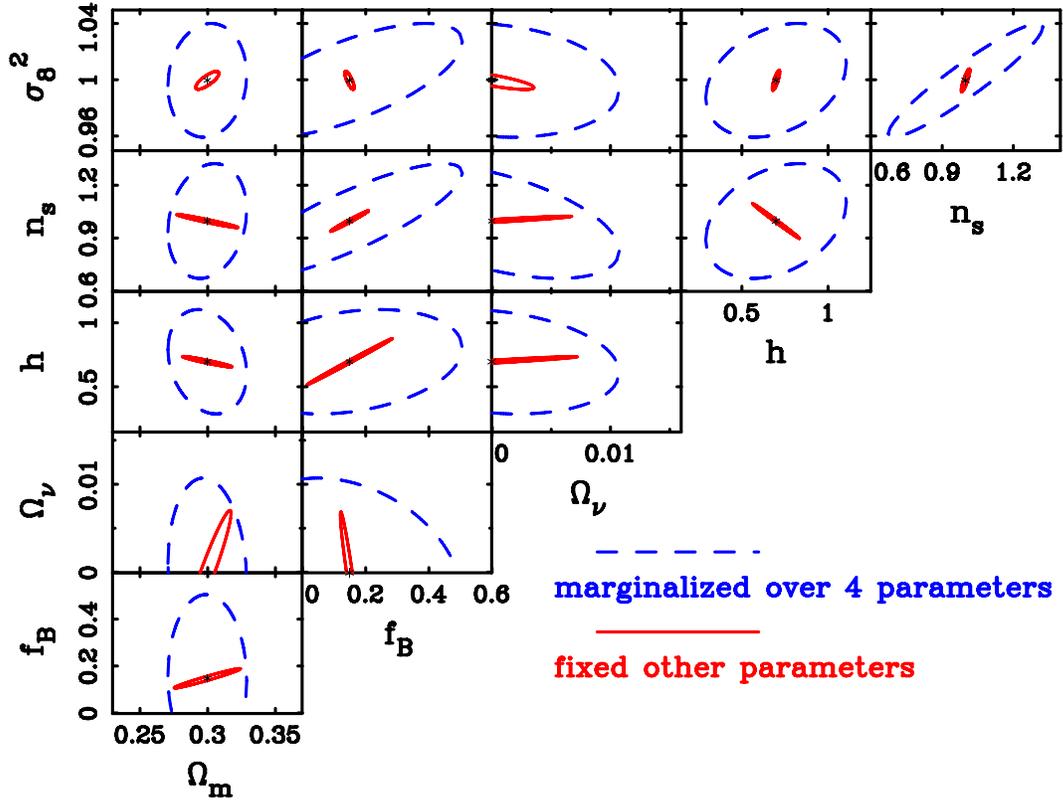}
  \end{center}
  \caption{The minimum variance bounds of the parameters expected from
    the LRG sample. We use the Epanechnikov kernel, i.e. $m=1$. A
    number of data points is six hundred. {\it Solid lines}:
    1-$\sigma$ confidence level regions when we carry out the
    two-parameter estimation, showing how two parameters are
    degenerate each other; {\it dashed lines}: 1-$\sigma$ confidence
    level regions when we fit all six parameters at a
    time. \label{fig:fig3}}
\end{figure}

\begin{figure}[h]
  \begin{center}
    \FigureFile(140mm,100mm){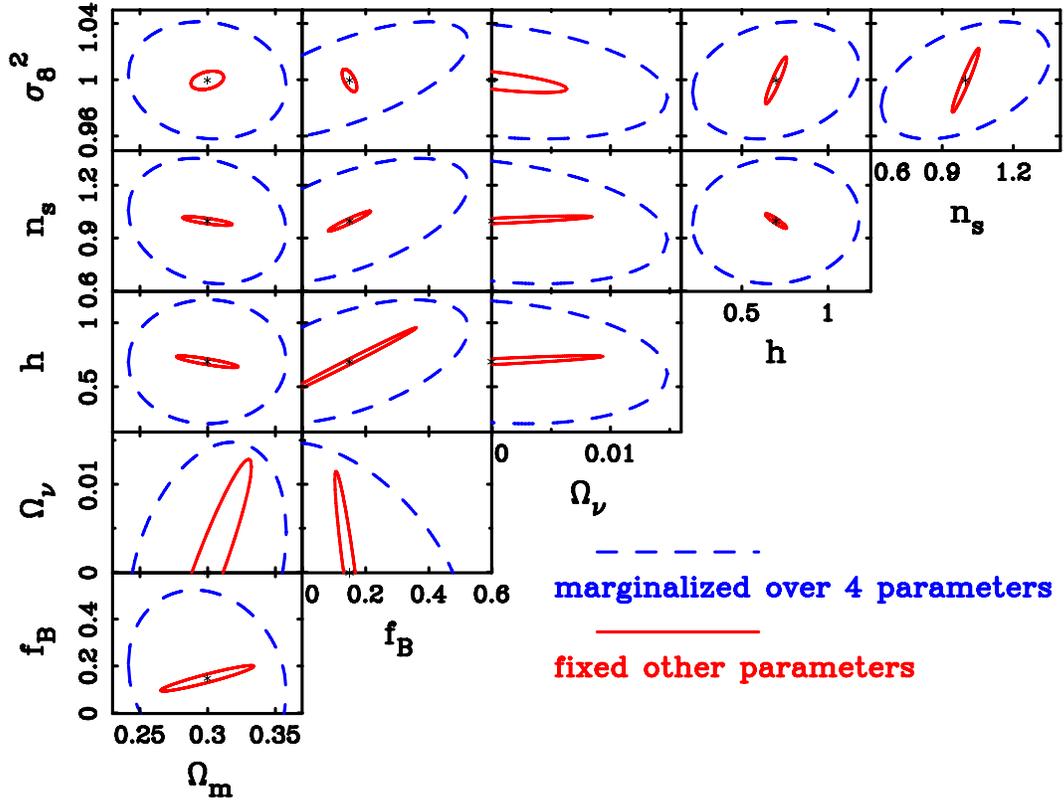}
  \end{center}
  \caption{The minimum variance bounds of the parameters expected from
    the MG sample. We use the Epanechnikov kernel, i.e. $m=1$. A
    number of data points is four hundred. Lines have same meanings as
    figure~\ref{fig:fig3}. \label{fig:fig4}}
\end{figure}

Figure~\ref{fig:fig3} and Figure~\ref{fig:fig4} demonstrate expected
parameter bounds for the LRG and for the MG, respectively. In both
Figures, solid lines represent $1\sigma$ confidence ellipsoids when we
carry out two-parameter likelihood analysis fixing all the other
parameters. Dashed lines show the $1\sigma$ confidence ellipsoids when
we carry out likelihood analysis for the all parameters at a time and
marginalized over the all parameters but two. In all cases, the LRG
can constrain the parameters more tightly than the MG, because the
survey volume is larger in LRG.

\begin{figure}[h]
  \begin{center}
    \FigureFile(105mm,75mm){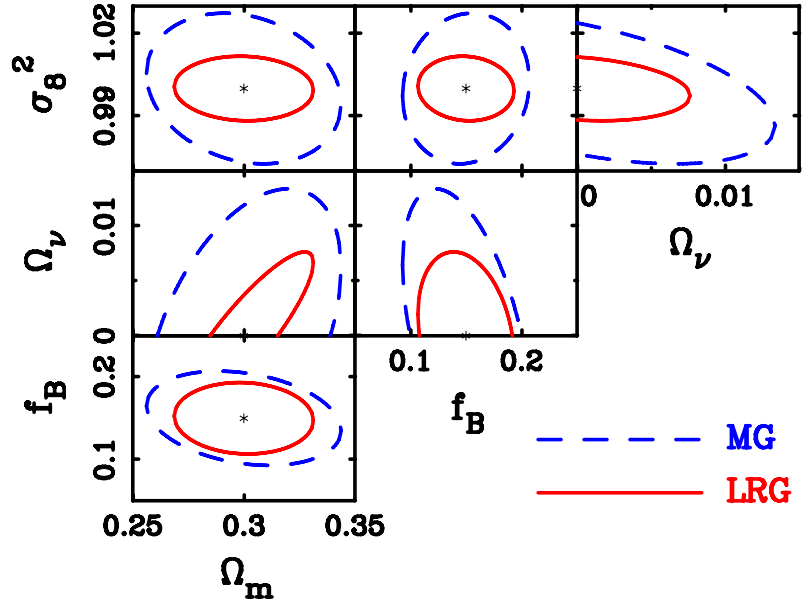}
  \end{center}
  \caption{Confidence contours marginalized over four parameters,
    i.e. the matter density parameter, the baryon fraction, the
    neutrino density parameter and ${\sigma_8}^2$. {\it Solid lines}:
    1-$\sigma$ confidence ellipsoids for the LRG sample; {\it dashed
      lines}: 1-$\sigma$ confidence ellipsoids for the MG
    sample.\label{fig:fig5}}
\end{figure}

The expected parameter bounds in the case all the other parameters are
marginalized over are much broader than in the case of just fitting
only two parameters. Parameter bounds for $f_{\rm B}$, $h$, and
$n_{\rm s}$ are especially broadened. This is mainly because there are
degeneracies among these parameters, which one can notice in the
Figures~\ref{fig:fig3} and \ref{fig:fig4}. Therefore, simultaneous
determination of all the parameters considered here is not efficient
when only the second moments are used. It is rather preferable to
consider that such degenerate parameters are sufficiently constrained
by other observations. Figure~\ref{fig:fig5} shows the parameter
bounds in the case that $h$ and $n_{\rm s}$ are fixed. Dashed lines
are $1\sigma$ confidence ellipsoids for the MG. The solid lines are
for the LRG. One sees that the expected bounds are much tighter than
previous figures, and degeneracies between parameters are resolved.

\section{Conclusion and Discussion}
\label{sec:discussion}

We consider the CIC analysis of the large-scale galaxy distribution.
We estimate the expected parameter bounds from the second moments by
means of the Fisher matrix analysis, considering SDSS-like surveys as
working examples. In this analysis, we investigate a series of
smoothing functions, expecting that there is an optimal smoothing
function with respect to the parameter bounds. We derive an analytic
form of the covariance matrix. As a result, we found that it is
possible to improve the parameter bounds by using an optimized kernel;
the Epanechnikov kernel gives a better parameter bounds than
traditional top-hat and Gaussian kernels. As previous study shows, the
LRG sample is more suitable than the MG sample for parameter
estimation because of the larger observational volume.

We carried out two other tests to investigate the reason why the
Epanechnikov kernel gives the tightest constraints among all
generalized Epanechnikov kernels. First, we test the hypothesis that
the oscillating phases in the power spectrum and that in smoothing
kernels are synchronized with each other. It seems this hypothesis is
not realistic because of the comparison in Figure~\ref{fig:nobaryon}.
Actually, we do not find clear evidence of the phase synchronization,
directly comparing the both oscillations.

Second, we test the hypothesis that the signal to noise ratio
significantly depends on the shape of the kernels. As a result, we do
not find signficant difference between the top-hat smoothing function
and the Epanechnikov kernels. Therefore, we have not identify a clear
reason why the Epanechnikov kernel is the best, which remains to be
studied in the future work.

From our results, the following points are notable for parameter
estimation. About 400 and 600 different smoothing scales are
sufficient to minimize the parameter bounds for the MG sample and LRG
sample, respectively. Due to the covariance between data with
different smoothing scales, too many data with different smoothing
radii have a redundant information. In addition, the parameters
$\Omega_{\rm B}$, $h$, and $n_{\rm s}$ degenerate in the parameter
estimation. This degeneracy is due to the fact that varying these
parameters similarly change the shape of the power spectrum on scales
of interest.

In this analysis, we do not explicitly consider redshift-space
distortion effects. In treating an actual galaxy distributions, radial
distances are measured by redshifts, and the redshift-space clustering
of galaxies is the observable \citep{reddist}. In
equations~(\ref{eq:B21}) and (\ref{eq:B22}), integrations over
$\boldsymbol{r}_\alpha$ and $\boldsymbol{r}_\beta$ do not have
anything to do with a line-of-sight direction. Thus, integrations of
equations~(\ref{eq:same}) and (\ref{eq:diff}) results in
equations~(\ref{eq:xibarsame}) and (\ref{eq:xibardiff}). In addition,
we can take $z$-axis as a line-of-sight direction in integrating over
$\boldsymbol{k}$. Therefore, we find that this modifies the covariance
matrix in such a way that an additional factor is added to the cosmic
variance. As a result, the last terms of equations~(\ref{eq:B21})
and (\ref{eq:B22}) are multiplied by a factor $\left(1 + 2\beta/3 +
  \beta^2/5\right)$, where $\beta$ is a redshift-space distortion
factor and is approximately given by $\beta=\Omega_{\rm m}^{0.6}/b$,
and $b$ is the linear bias factor.  Therefore, the redshift-space
effects are absorbed in the normalization of the power spectrum.

In Appendix~\ref{Ap:moments-counts}, we derive relations between
higher-order moments, and galaxy counts for an arbitrary smoothing
function. Although we analyze only second moment in this paper,
higher-order moments provide additional information on cosmology. One
can expect the possibility to search for an optimal smoothing function
for the higher-order moments in CIC analysis. By combining the second
moment and higher-order moments, one can break the degeneracy between
the bias $b$ and the normalization ${\sigma_8}^2$ and reduce other
degeneracies in parameters \citep{bispectrum}. Theoretical predictions
for the higher-order moments can be found in
\citet{skewness,kurtosis}. In a case involving bias parameter,
relations between reduced moments of galaxies and that of matter are
given in \citet{bias}. However, analytical treatment of the covariance
matrix for higher-order moments will be more involving. In that case,
some numerical evaluation of the covariances would be a realistic
method.

In the case of evaluating the power spectrum $P(k)$ of the 2dFGRS, 10
simulated realizations are used to take into account non-linear
effects, redshift-space effects, and geometric effects
\citep{2df-para}. To obtain an accurate covariance matrix of $P(k)$,
they need about 1000 realizations. It is not realistic to run full
N-body simulations, and they use realizations of random Gaussian
fields for that purpose. Otherwise, one can use the PTHALOS code to
generate the mock galaxy distribution that are based on the
conjunction of a halo model and a second-order perturbation theory
\citep{pthalos}. In addition, when we carry out the likelihood
analysis with many parameters, the grid-based likelihood analysis is
also unrealistic because of the cost of CPU time. Recently many teams
use the MCMC method \citep{mcmc,wmap-mcmc} to obtain the maximum
likelihood of parameters in a reasonable time scale.

\bigskip
We wish to thank Seiko Inoue for her help in earlier stages of this
work. We acknowledge the CMBfast code which is publicly available.
T.M. acknowledges support from the JSPS Grant-in-Aid for Scientific
Research, 18540260, 2006.

\appendix
\section{Relations Between Reduced Moments and Galaxy Counts with an
  Arbitrary Smoothing Function}
\label{Ap:moments-counts}

\subsection{The second moment}
\label{Ap:variance}

\begin{figure}[h]
  \begin{center}
    \FigureFile(70mm,50mm){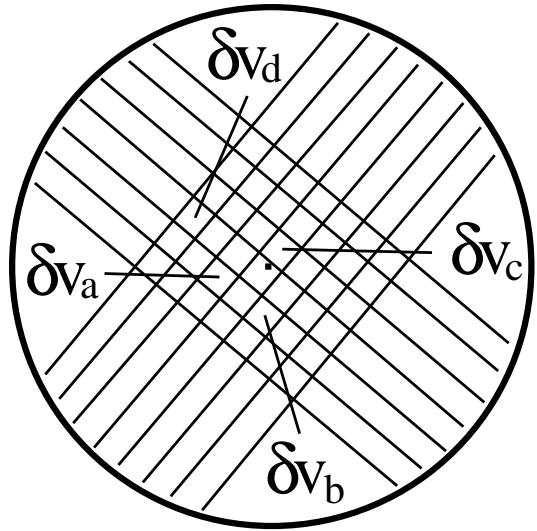}
  \end{center}
  \caption{Schematic picture of dividing cells into a series of
    infinitesimal volumes. \label{fig:fig8}}
\end{figure}

In this section, we obtain a expression of a second moment of density
contrast in terms of statistics of numbers of galaxies in a cell,
following a similar procedure given by \citet{peebles}. First, we
divide the cells into a series of infinitesimal volumes, $\delta v_a$,
as shown in Figure~\ref{fig:fig8}. A number of galaxies, $n_a$, in an
infinitesimal volume $\delta v_a$ is either 0 or 1.  Therefore,
\begin{equation}
  \langle n_a \rangle =
  \left\langle {n_a}^2 \right\rangle =
  \left\langle {n_a}^3 \right\rangle =
  \cdots = \overline{n}\delta v_a,
\end{equation}
where $\overline{n}$ is a mean number density of galaxies in the
universe.  The count of galaxies $N$ defined by equation~(\ref{eq:40})
is given by
\begin{equation}
  N = \sum_a n_a W\left(|\boldsymbol{r}_a|; R\right),
\label{eq:count}
\end{equation}
where the origin of coordinates is taken at the center of the cell
$\boldsymbol{r}_{\rm a}$. The mean value of the count is given by
\begin{equation}
  \langle N \rangle = 
  \sum_a \left\langle n_a \right\rangle W\left(|\boldsymbol{r}_a|; R\right) =
  \overline{n} \int d^3r_a \, W\left(|\boldsymbol{r}_a|; R\right) =
  \overline{n}v,
\end{equation}
where $v=4\pi R^3/3$. Similarly, a mean square of $N$ is calculated as
\begin{eqnarray}
  && \hspace{-1em} \left\langle N^2 \right\rangle
  \nonumber \\
  &=&
  \left\langle
    \sum_a n_a W\left(|\boldsymbol{r}_a|,R\right) \cdot
    \sum_b n_b W\left(|\boldsymbol{r}_b|;R\right)
  \right\rangle
  \nonumber\\
  &=&
  \sum_a \left\langle {n_a}^2 \right\rangle
  \left\{ W\left(|\boldsymbol{r}_a|; R\right) \right\}^2 +
  \sum_{a\neq b} \left\langle n_an_b \right\rangle
  W\left(|\boldsymbol{r}_a|; R\right)W\left(|\boldsymbol{r}_b|; R\right)
  \nonumber\\
  &=&
  \overline{n}
  \int d^3r_a \, \left\{ W\left(|\boldsymbol{r}_a|;R\right) \right\}^2 + 
  \overline{n}^2 \int d^3r_ad^3r_b \, 
  \left\{
    1 + \xi\left(|\boldsymbol{r}_a-\boldsymbol{r}_b|\right)
  \right\}
  W(|\boldsymbol{r}_a|; R)W(|\boldsymbol{r}_b|; R)
  \nonumber\\
  &=&
  \langle N \rangle
  \overline{W} +
  \langle N \rangle^2 
  \left(
    1 + \frac{1}{v^2}
    \int d^3r_ad^3r_b \,
    \xi(|\boldsymbol{r}_a-\boldsymbol{r}_b|)
    W(|\boldsymbol{r}_a|; R)
    W(|\boldsymbol{r}_b|; R)
  \right).
\end{eqnarray}
Therefore, we obtain a relation between the number of galaxies and the
second moment by using equation~(\ref{eq:variance})
\begin{equation}
  {\sigma_R}^2 = \frac{\langle N^2\rangle}{\langle N\rangle^2} 
  - \frac{\overline{W}}{\langle N\rangle}-1, \label{eq:sigma}
\end{equation}
where $\overline{W}$ is given by equation~(\ref{eq:60}). Below, we
need a generalized quantity $\overline{W^n}$ which is defined by
\begin{equation}
  \overline{W^n} =
  \frac{3}{4\pi R^3} \int d^3r \, \left\{W(|\boldsymbol{r}|;R)\right\}^{n+1}
\label{eq:a10}
\end{equation}
In the case of $m$-weight Epanechnikov kernel, this quantity is
analytically calculated and is given by
\begin{equation}
  \overline{\left(W_m\right)^n} =
  \frac{3\sqrt{{\pi}}}{4}
  \frac{\Gamma(m(n+1)+1)}{\Gamma(m(n+1)+5/2)}
  \left(\frac{4}{3\sqrt{\pi}}\frac{\Gamma(m+5/2)}{\Gamma(m+1)}\right)^{n+1},
\label{eq:v-ratio}
\end{equation}
where $\Gamma(x)$ is the gamma function. For the top-hat smoothing
function ($m=0$), the quantity $\overline{W^n}$ is simply unity for
any $n$.

\subsection{The Skewness}
A similar relation for the third moments, or its reduced quantity
skewness, can be found. In calculating $\left\langle
  N^3\right\rangle$, there exist a term with a weight of
$\{W(r)\}^2$. We need to define another kind of galaxy count by
\begin{equation}
  N^{(p+1)} = \sum_a n_a \left\{ W(|\boldsymbol{r}_a|; R) \right\}^{p+1},
\end{equation}
:which corresponds to a weighted count with a smoothing function
$\{W(r)\}^{p+1}$. The mean value of this count is given by
\begin{equation}
  \left\langle N^{(p+1)} \right\rangle =
  \sum_a\langle n_a\rangle
  \left\{ W(|\boldsymbol{r}_a|; R) \right\}^{p+1} = 
  \overline{n}\int d^3r_a \, \left\{ W(|\boldsymbol{r}_a|; R) \right\}^{p+1} =
  \langle N\rangle\overline{W^p}.
\end{equation}
In the following, we also need a quantity,
\begin{eqnarray}
  \left\langle N^2 \right\rangle^{(p,q)}
  &\equiv&
  \left\langle N^{(p+1)} \cdot N^{(q+1)} \right\rangle
  \nonumber\\
  &=&
  \overline{n} \int d^3r_a \,
  \left\{ W(|\boldsymbol{r}_a|; R) \right\}^{p+q+2}
  \nonumber \\
  && +
  \overline{n}^2\int d^3r_ad^3r_b\,(1+\xi_{ab})
  \left\{ W(|\boldsymbol{r}_a|; R) \right\}^{p+1}
  \left\{ W(|\boldsymbol{r}_b|; R) \right\}^{q+1}
  \nonumber\\
  &=&
  \langle N\rangle\overline{W^{p+q+1}}+\langle
  N\rangle^2\overline{W^p}\,
  \overline{W^q}\left(1+\overline{\xi}_R^{(p,q)}\right)
\label{eq:2nd-cross-moment},
\end{eqnarray}
where $\xi_{ab}=\xi(|\boldsymbol{r}_a-\boldsymbol{r}_b|)$ is the
two-point correlation function and we define
\begin{equation}
  \overline{\xi}_R^{(p,q)} \equiv
  \frac{\displaystyle
    \int d^3r_ad^3r_b\,\xi(|\boldsymbol{r}_a-\boldsymbol{r}_b|)
    \left\{ W(|\boldsymbol{r}_a|; R) \right\}^{p+1}
    \left\{ W(|\boldsymbol{r}_b|; R) \right\}^{q+1}
  }
  {\displaystyle
    \int d^3r_a \, \left\{ W(|\boldsymbol{r}_a|; R) \right\}^{p+1}
    \int d^3r_b \, \left\{ W(|\boldsymbol{r}_b|; R) \right\}^{q+1}
  }.
\label{eq:same}
\end{equation}
Note that $\overline{\xi}_R^{(0,0)} = \left\langle {\delta_R}^2
\right\rangle = \sigma^2(R)$. We are ready to calculate $\left\langle
  N^3 \right\rangle$ for an arbitrary smoothing function. It is given
by
\begin{eqnarray}
  \langle N^3 \rangle
  &=&
  \left\langle
    \sum_a n_a W(|\boldsymbol{r}_a|; R) \cdot
    \sum_b n_b W(|\boldsymbol{r}_b|; R) \cdot
    \sum_c n_c W(|\boldsymbol{r}_c|; R)
  \right\rangle
  \nonumber\\
  &=&
  \langle N\rangle\overline{W^2} +
  3\langle N\rangle^2\overline{W}
  \left(1+\overline{\xi}_R^{(1,0)}\right) +
  \langle N\rangle^3\left(1+3\overline{\xi}_R^{(0,0)} +
    \overline{\zeta}_R\right),
\label{eq:3rd-moment}
\end{eqnarray}
where $\overline{\zeta}_R$ is an ensemble average of the third power
of a smoothed field and corresponds to a volume average of the
three-point correlation function,
$\zeta(\boldsymbol{r}_a,\boldsymbol{r}_b,\boldsymbol{r}_c)$:
\begin{eqnarray}
  \overline{\zeta}_R &\equiv&
  \frac{1}{v^3} \int_V d^3r_ad^3r_bd^3r_c \,
  \zeta(\boldsymbol{r}_a,\boldsymbol{r}_b,\boldsymbol{r}_c)
  W(|\boldsymbol{r}_a|;R) W(|\boldsymbol{r}_b|;R) W(|\boldsymbol{r}_c|;R)
  \nonumber \\
  &=&
  \left\langle {\delta_R}^3 \right\rangle.
\label{eq:averaged-3pt}
\end{eqnarray}
The skewness parameter is defined by
\begin{equation}
  S_3 \equiv
  \frac{\left\langle {\delta_R}^3\right\rangle}
    {\left\langle {\delta_R}^2\right\rangle^2}.
\end{equation}
By using equations~(\ref{eq:2nd-cross-moment}) -
(\ref{eq:averaged-3pt}), we obtain
\begin{equation}
  S_3 = 
  \frac{\displaystyle
    \left\langle N\right\rangle
    \left(
      \left\langle N^3\right\rangle -
      3\left\langle N\right\rangle\left\langle N^2\right\rangle -
      3\left\langle N^2\right\rangle^{(1,0)} +
      3\left\langle N\right\rangle^2\overline{W} +
      2\left\langle N\right\rangle\overline{W} +
      2\langle N\rangle^3
    \right)  }
  {\displaystyle \left(\langle N^2\rangle -
      \langle N\rangle\overline{W} -
      \langle N\rangle^2\right)^2}.
\end{equation}

\subsection{The Kurtosis}
The fourth moment, or its reduced quantity kurtosis, is similarly
calculated as above. We define
\begin{eqnarray}
  && \hspace{-1em} \left\langle N^3 \right\rangle^{(p,q,r)}
  \nonumber \\
  &\equiv&
  \left\langle N^{(p+1)} \cdot N^{(q+1)} \cdot N^{(r+1)} \right\rangle
\nonumber\\
  &=&
  \langle N\rangle\overline{W^{p+q+r+2}} +
  \langle N\rangle^2\,\overline{W^{p+q+1}}\,
  \overline{W^r}\left(1+\overline{\xi}_R^{(p+q+1,r)}\right)
\nonumber\\
  && +
  \langle N\rangle^2\overline{W^{p+r+1}}\,\overline{W^q}
  \left(1+\overline{\xi}_R^{(p+r+1,q)}\right) +
  \langle N\rangle^2\,\overline{W^p}\,\overline{W^{q+r+1}}
  \left(1+\overline{\xi}_R^{(p,q+r+1)}\right)
\nonumber\\
  && +
  \langle N\rangle^3\,\overline{W^p}\,\overline{W^q}\,\overline{W^r}
  \left(1+\overline{\xi}_R^{(p,q)} +
    \overline{\xi}_R^{(p,r)} +
    \overline{\xi}_R^{(q,r)} +
    \overline{\zeta}_R^{(p,q,r)}
  \right),
\label{eq:3rd-cross-moment}
\end{eqnarray}
where $\overline{\zeta}_R^{(p,q,r)}$ is defined by
\begin{eqnarray}
&&\overline{\zeta}_R^{(p,q,r)} \equiv\nonumber\\
  &&  \int d^3r_ad^3r_bd^3r_c \, 
    \zeta(\boldsymbol{r}_a,\boldsymbol{r}_b,\boldsymbol{r}_c)
    \{ W(|\boldsymbol{r}_a|; R) \}^{p+1} 
    \{ W(|\boldsymbol{r}_b|; R) \}^{q+1}
    \{ W(|\boldsymbol{r}_c|; R) \}^{r+1}
  \nonumber \\
  && \times
  \left[
    \int d^3r_a \, 
    \{ W(|\boldsymbol{r}_a|; R) \}^{p+1}
    \int d^3r_b \, 
    \{ W(|\boldsymbol{r}_b|; R) \}^{q+1}
    \int d^3r_c \, 
    \{ W(|\boldsymbol{r}_c|; R) \}^{r+1}
  \right]^{-1}.
  \label{eq:zeta}
\end{eqnarray}
The fourth moment of galaxy count $\left\langle N^4 \right\rangle$ is
obtained as
\begin{eqnarray}
  \left\langle N^4 \right\rangle
  &=&
  \langle N \rangle \overline{W^3} +
  4 \langle N \rangle^2 \overline{W^2}
  \left(
    1 + \overline{\xi}_R^{(2,0)}
  \right)
  \nonumber\\
  && + \,
  3 \langle N \rangle^2
  \left( \overline{W} \right)^2
  \left(
    1 + \overline{\xi}_R^{(1,1)}
  \right) +
  6 \langle N \rangle^3 \, \overline{W}
  \left(
    1 + 2\overline{\xi}_R^{(1,0)} +
    {\sigma_R}^2+\overline{\zeta}_R^{(1,0,0)}
  \right)
  \nonumber\\
  && +\,
  \langle N \rangle^4
  \left(
    1 + 6{\sigma_R}^2 + 4\overline{\zeta}_R + 3 {\sigma_R}^4 + \overline{\eta}_R
  \right),
\end{eqnarray}
where $\overline{\eta}_R$ is a volume average of the four-point
correlation function,
$\eta(\boldsymbol{r}_a,\boldsymbol{r}_b,\boldsymbol{r}_c,\boldsymbol{r}_d)$,
and is given by
\begin{eqnarray}
  \overline{\eta}_R
  &\equiv&
  \frac{1}{v^4}\int d^3r_ad^3r_bd^3r_cd^3r_d\,
  \eta(\boldsymbol{r}_a,\boldsymbol{r}_b,\boldsymbol{r}_c,\boldsymbol{r}_d)
  W(|\boldsymbol{r}_a|; R)W(|\boldsymbol{r}_b|; R)
  W(|\boldsymbol{r}_c|; R)W(|\boldsymbol{r}_d|; R)
  \nonumber\\
  &=& \left\langle {\delta_R}^4 \right\rangle.
  \label{eq:averaged-4pt}
\end{eqnarray}
The reduced kurtosis is defined by
\begin{equation}
  S_4 \equiv
  \frac{\left\langle{\delta_R}^4\right\rangle -
    3\left\langle{\delta_R}^2\right\rangle^2}
  {\left\langle{\delta_R}^2\right\rangle^3}.
\end{equation}
Therefore, by using equations~(\ref{eq:2nd-cross-moment}) and
(\ref{eq:3rd-cross-moment}) - (\ref{eq:averaged-4pt}), we obtain
\begin{eqnarray}
  S_4 &=&
  \left\langle N \right\rangle^2
  \left[
    \left\langle N^4 \right\rangle -
    4 \left\langle N \right\rangle \left\langle N^3 \right\rangle +
    18 \left\langle N \right\rangle^2 \left\langle N^2 \right\rangle -
    6 \left\langle N^2 \right\rangle^2 -
    9 \left\langle N \right\rangle^4 -
    6 \left\langle N^3 \right\rangle^{(1,0,0)}
  \right.
  \nonumber\\
  && +
    12 \left\langle N \right\rangle
    \left\langle N^2 \right\rangle^{(1,0)} +
    12 \overline{W} \left\langle N \right\rangle 
    \left\langle N^2\right\rangle -
    18 \overline{W} \left\langle N \right\rangle^3 +
    8 \left\langle N^2 \right\rangle^{(2,0)} +
    3 \left\langle N^2 \right\rangle^{(1,1)}
  \nonumber\\
  && -
  \left.
    8 \overline{W^2} \left\langle N \right\rangle^2 -
    6 \left( \overline{W} \right)^2 \left\langle N \right\rangle^2 -
    6 \overline{W^3} \left\langle N \right\rangle
  \right]
    \left[
      \left\langle N^2 \right\rangle -
      \left\langle N \right\rangle \overline{W} -
      \left\langle N \right\rangle^2
    \right]^{-3}
\end{eqnarray}

\section{Derivation of the Covariance Matrix}
\label{Ap:covariance-matrix}

In this Appendix, we derive an analytic expression of the covariance
matrix. Some complication arises to take overlappings of cells into
account. We assume a locally Poisson process in the overlapping
regions for simplicity, following \citet{over,lap}. The definition of
the covariance matrix is given by equation~(\ref{eq:160}). In the
following, we denote the galaxy count with a smoothing radius $R_i$ by
$N_i$. Individual cells are labeled by $\alpha, \beta$ and
$N_{i,\alpha}$ is a galaxy count in a cell $\alpha$ which has a
smoothing radius $R_i$. A label $\alpha$ is used for a group cells
having radii $R_i$ and a label $\beta$ is for group of cells having
radii $R_j$.

The second moment of the density contrast is related to galaxy counts
by equation~(\ref{eq:50}). The mean square of the galaxy count is
given by
\begin{equation}
  \left\langle {N_i}^2 \right\rangle = 
  \frac{1}{N_{{\rm cell},i}}
  \sum_{\alpha=1}^{N_{{\rm cell},i}}
  {N_{i,\alpha}}^2,
\label{eq:b10}
\end{equation}
where $N_{{\rm cell},i}$ is the number of cells of radius $R_i$. We
assume that the mean galaxy count $\langle N_i \rangle$ is accurately
approximated by $\overline{n}v_i$, where $v_i = 4\pi {R_i}^3/3$, and
that the uncertainty of the mean number density $\overline{n}$ is
sufficiently small. This assumption corresponds to the case when we
know a precise selection function from another observation and a
sufficiently accurate mean number density from that selection
function. Therefore, if we estimate mean numbers of galaxies from a
same data set, the above assumption is inaccurate. By combining
equations~(\ref{eq:50}), (\ref{eq:160}), and (\ref{eq:b10}), the
expression of the covariance matrix is obtained. After some
calculation, we have
\begin{eqnarray}
  C_{ii} 
  &=&
  \frac{1}{\left\langle N_i \right\rangle^4}
  \left[
    \frac{1}{{N_{{\rm cell},i}}^2}
    \sum^{N_{{\rm cell},i}}_\alpha
    \left\langle {N_{i,\alpha}}^4 \right\rangle
  \right.
  \nonumber \\
  &&\hspace{5em} +
  \left.
    \left(1-\frac{1}{N_{{\rm cell},i}}\right)
    \frac{1}{V^2}\int_V d^3r_\alpha d^3r_\beta\,
    \left\langle {N_{i,\alpha}}^2 {N_{i,\beta}}^2 \right\rangle -
    \left\langle N_i^2 \right\rangle^2
  \right],
  \label{eq:B2} \\
  C_{ij}
  &=&
  \frac{1}{\left\langle N_i \right\rangle^2 \left\langle N_j \right\rangle^2}
  \left[
    \frac{1}{V^2} \int_V d^3r_\alpha d^3r_\beta \,
    \left\langle {N_{i,\alpha}}^2 {N_{j,\beta}}^2 \right\rangle -
    \left\langle N_i^2 \right\rangle \left\langle N_j^2 \right\rangle
  \right],
\label{eq:B3}
\end{eqnarray}
where $V$ indicates a survey volume and $\left\langle {N_{i,\alpha}}^2
    {N_{j,\beta}}^2 \right\rangle$ is a product's mean value of the
squared galaxy counts in two different cells. In this study, we
consider that cells are located randomly in a spherical survey volume
by following \citet{over,lap}. Then, the sums over cells can be
replaced by integrations over survey volume, as $\sum_\alpha (\cdots)
= N_{{\rm cell},i} /V \int_V d^3r_\alpha\, (\cdots)$. However, in
equation~(\ref{eq:B2}), we can divide $\left\langle {N_{i,\alpha}}^2
    {N_{i,\beta}}^2 \right\rangle$ into two cases depending on
$\alpha=\beta$ or not. $\left\langle {N_{i,\alpha}}^4 \right\rangle$
is a term that comes from a case of $\alpha=\beta$ and represents an
averaged value of biquadratic weighted counts over all cells. The
second term comes from a case where $\alpha\neq\beta$ holds. An
additional factor $(1 - 1/N_{{\rm cell},i})$ in front of this term
should be present because of the following reason: The number of cases
satisfying $\alpha\neq\beta$ is ${N_{{\rm cell},i}}^2 - N_{{\rm
    cell},i}$ Therefore, the sum is replaced by the volume integration
as $\sum_{\alpha\neq\beta}\sum_\beta(\cdots) = \left({N_{{\rm
          cell},i}}^2 - N_{{\rm cell},i} \right) /V^2 \int d^3r_\alpha
d^3r_\beta\,(\cdots)$. $\left\langle {N_i}^2 \right\rangle$ is a
theoretically expected square mean of number count related with the
variance by equation~(\ref{eq:sigma}).

To calculate each term, we define a similar notation as equation
(\ref{eq:same}):
\begin{equation}
  \overline{\xi}_{\alpha\beta}^{(p,q)}
  \equiv 
  \frac{\displaystyle
    \int d^3r_ad^3r_b \, 
    \{ W(|\boldsymbol{r}_a-\boldsymbol{r}_\alpha|; R_i) \}^{p+1} 
    \{ W(|\boldsymbol{r}_b-\boldsymbol{r}_\beta|; R_j) \}^{q+1}
    \xi_{ab}
  }
  {\displaystyle 
    \int d^3r_a \, 
    \{ W(|\boldsymbol{r}_a-\boldsymbol{r}_\alpha|; R_i) \}^{p+1}
    \int d^3r_b \, 
    \{ W(|\boldsymbol{r}_b-\boldsymbol{r}_\beta|; R_j) \}^{q+1}
  },
  \label{eq:diff}
\end{equation}
where $\xi_{ab}=\xi(|\boldsymbol{r}_a-\boldsymbol{r}_b|)$ is the
two-point correlation function and $\boldsymbol{r}_\alpha$ and
$\boldsymbol{r}_\beta$ are centers of two
cells. Equation~(\ref{eq:same}) considers
correlations in a same cell. On the other hand, equation
(\ref{eq:diff}) considers correlations in two different cells. We also
use similar notations for the three- and four-point correlation
functions. In a case considering correlations in a same cell, we use
equation~(\ref{eq:zeta}) and
\begin{eqnarray}
  \overline{\eta}_{R}^{(p,q,r,s)} &\equiv&
  \int d^3r_ad^3r_bd^3r_cd^3r_d \, 
  \eta_{abcd}
  \{ W(|\boldsymbol{r}_a|; R) \}^{p+1} 
  \{ W(|\boldsymbol{r}_b|; R) \}^{q+1}
  \nonumber \\
  && \hspace{1em} \times
  \{ W(|\boldsymbol{r}_c|; R) \}^{r+1}
  \{ W(|\boldsymbol{r}_d|; R) \}^{s+1}
  \nonumber \\
  && \times
  \left[
    \int d^3r_a \, \{ W(|\boldsymbol{r}_a|; R) \}^{p+1}
    \int d^3r_b \, \{ W(|\boldsymbol{r}_b|; R) \}^{q+1}
  \right.
  \nonumber \\
  && \hspace{1em} \times
  \left.
    \int d^3r_c \, \{ W(|\boldsymbol{r}_c|; R) \}^{r+1}
    \int d^3r_d \, \{ W(|\boldsymbol{r}_d|; R) \}^{s+1}
  \right]^{-1},
  \label{eq:eta}
\end{eqnarray}
where we abbreviate the four-point correlation function
$\eta(\boldsymbol{r}_a,\boldsymbol{r}_b,\boldsymbol{r}_c,\boldsymbol{r}_d)$
as $\eta_{abcd}$. To consider correlations in two different cells
whose centers are located at $\boldsymbol{r}_\alpha,
\,\boldsymbol{r}_\beta$, we use
\begin{eqnarray}
\overline{\zeta}_{\alpha\alpha\beta}^{(p,q,r)} &\equiv&
    \int d^3r_ad^3r_bd^3r_c \, 
    \zeta_{abc}
    \{ W(|\boldsymbol{r}_a-\boldsymbol{r}_\alpha|; R_i) \}^{p+1} 
    \{ W(|\boldsymbol{r}_b-\boldsymbol{r}_\alpha|; R_i) \}^{q+1}
  \nonumber \\
  && \hspace{1em} \times
    \{ W(|\boldsymbol{r}_c-\boldsymbol{r}_\beta|; R_j) \}^{r+1}
  \nonumber \\
  && \times
  \left[
    \int d^3r_a \, 
    \{ W(|\boldsymbol{r}_a-\boldsymbol{r}_\alpha|; R_i) \}^{p+1}
    \int d^3r_b \, 
    \{ W(|\boldsymbol{r}_b-\boldsymbol{r}_\alpha|; R_j) \}^{q+1}
  \right.
  \nonumber \\
  && \hspace{1em} \times
  \left.
    \int d^3r_c \, 
    \{ W(|\boldsymbol{r}_c-\boldsymbol{r}_\beta|; R_i) \}^{r+1}
  \right]^{-1},\\
\overline{\zeta}_{\alpha\beta\beta}^{(p,r,s)} &\equiv&
    \int d^3r_ad^3r_cd^3r_d \, 
    \zeta_{acd}
    \{ W(|\boldsymbol{r}_a-\boldsymbol{r}_\alpha|; R_i) \}^{p+1} 
    \{ W(|\boldsymbol{r}_c-\boldsymbol{r}_\beta|; R_j) \}^{r+1}
  \nonumber \\
  && \hspace{1em} \times
    \{ W(|\boldsymbol{r}_d-\boldsymbol{r}_\beta|; R_j) \}^{s+1}
  \nonumber \\
  && \times
  \left[
    \int d^3r_a \, 
    \{ W(|\boldsymbol{r}_a-\boldsymbol{r}_\alpha|; R_i) \}^{p+1}
    \int d^3r_c \, 
    \{ W(|\boldsymbol{r}_c-\boldsymbol{r}_\beta|; R_i) \}^{r+1}
  \right.
  \nonumber \\
  && \hspace{1em} \times
  \left.
    \int d^3r_d \, 
    \{ W(|\boldsymbol{r}_d-\boldsymbol{r}_\beta|; R_j) \}^{s+1}
  \right]^{-1},\\
\overline{\eta}_{\alpha\alpha\beta\beta}^{(p,q,r,s)} &\equiv&
    \int d^3r_ad^3r_bd^3r_cd^3r_d \, 
    \eta_{abcd}
    \{ W(|\boldsymbol{r}_a-\boldsymbol{r}_\alpha|; R_i) \}^{p+1} 
    \{ W(|\boldsymbol{r}_b-\boldsymbol{r}_\alpha|; R_i) \}^{q+1}
  \nonumber \\
  && \hspace{1em} \times
    \{ W(|\boldsymbol{r}_c-\boldsymbol{r}_\beta|; R_j) \}^{r+1}
    \{ W(|\boldsymbol{r}_d-\boldsymbol{r}_\beta|; R_j) \}^{s+1}
  \nonumber \\
  && \times
  \left[
    \int d^3r_a \, 
    \{ W(|\boldsymbol{r}_a-\boldsymbol{r}_\alpha|; R_i) \}^{p+1}
    \int d^3r_b \, 
    \{ W(|\boldsymbol{r}_b-\boldsymbol{r}_\alpha|; R_j) \}^{q+1}
  \right.
  \nonumber \\
  && \hspace{1em} \times
  \left.
    \int d^3r_c \, 
    \{ W(|\boldsymbol{r}_c-\boldsymbol{r}_\beta|; R_i) \}^{r+1}
    \int d^3r_d \, 
    \{ W(|\boldsymbol{r}_d-\boldsymbol{r}_\beta|; R_j) \}^{s+1}
  \right]^{-1},
  \label{eq:four}
\end{eqnarray}
where we abbreviate the three-point correlation function
$\zeta(\boldsymbol{r}_a,\boldsymbol{r}_b,\boldsymbol{r}_c)$ as
$\zeta_{abc}$.

The first term in equation~(\ref{eq:B2}) is calculated in the same
manner as in Appendix~\ref{Ap:moments-counts}. Then, we obtain
\begin{eqnarray}
  \left\langle {N_{i,\alpha}}^4 \right\rangle
  &=&
  \left\langle
    \left( \sum_{a\in\alpha} 
      n_a W(|\boldsymbol{r}_a-\boldsymbol{r}_\alpha|;R_i) \right)^4
  \right\rangle
  \nonumber \\
  &=&
  \langle N \rangle 
  \overline{W^3} + 
  4\langle N \rangle^2
  \overline{W^2} 
  \left(
    1 + \overline{\xi}_{R_i}^{(2,0)}
  \right) +
  3\langle N \rangle^2
  \left( \overline{W} \right)^2
  \left(
    1 + \overline{\xi}_{R_i}^{(1,1)}
  \right)
  \nonumber \\
  && +
  6 \langle N \rangle^3
  \overline{W}
  \left(
    1 + \overline{\xi}_{R_i}^{(0,0)} +
    2 \overline{\xi}_{R_i}^{(1,0)} + \zeta_{R_i}^{(1,0,0)}
  \right)
  \nonumber \\
  && +
  \langle N \rangle^4
  \left[
    1 + 6 \overline{\xi}_{R_i}^{(0,0)} +
    3 \left( \overline{\xi}_{R_i}^{(0,0)} \right)^2 +
    4 \zeta_{R_i}^{(0,0,0)} +
    \eta_{R_i}^{(0,0,0,0)}
  \right],
\label{eq:B8}
\end{eqnarray}
where $\sum_{a\in\alpha}$ represents a summation over every infinitesimal
volumes $\delta v_a$ in a cell $\alpha$.

From equation~(\ref{eq:sigma}), the last term in both equations
(\ref{eq:B2}), (\ref{eq:B3}) is given by
\begin{eqnarray}
  && \hspace{-1em} \left\langle {N_i}^2 \right\rangle
  \left\langle {N_j}^2 \right\rangle
  \nonumber \\
  &=& 
    \langle N_i \rangle
    \langle N_j \rangle 
    \left( \overline{W} \right)^2 +
    \langle N_i \rangle
    \langle N_j \rangle^2
    \overline{W}
    \left(
      1 + \overline{\xi}_{R_j}^{(0,0)}
    \right) +
    \langle N_i \rangle^2
    \langle N_j \rangle
    \overline{W}
    \left(
      1 + \overline{\xi}_{R_i}^{(0,0)}
    \right)
  \nonumber \\
  && +
    \langle N_i \rangle^2
    \langle N_j \rangle^2
    \left(
      1 + \overline{\xi}_{R_i}^{(0,0)} +
      \overline{\xi}_{R_j}^{(0,0)} +
      \overline{\xi}_{R_i}^{(0,0)} 
      \overline{\xi}_{R_j}^{(0,0)}
    \right)
\label{eq:B9}.
\end{eqnarray}

A derivation of the second term in equation~(\ref{eq:B2}) is
complicated because of overlappings of cells. First of all, the
integration is divided into two ranges of $r$ depending on whether
cells are overlapping or not, i.e. $\int (\cdots) d^3r_\alpha
d^3r_\beta = \int_{r_{\alpha\beta}\le R_i+R_j} (\cdots) d^3r_\alpha
d^3r_\beta + \int_{r_{\alpha\beta}>R_i+R_j} (\cdots) d^3r_\alpha
d^3r_\beta$, where $r_{\alpha\beta} = |\boldsymbol{r}_\alpha -
\boldsymbol{r}_\beta|$. On the one hand, for cell pairs that are not
overlapped to each other, we calculate $\left\langle {N_{i,\alpha}}^2
  {N_{j,\beta}}^2 \right\rangle$ in the same way as equation
(\ref{eq:B8}). Then, we obtain
\begin{eqnarray}
  && \hspace{-1em} \left\langle {N_{i,\alpha}}^2 {N_{j,\beta}}^2 \right\rangle
  \nonumber \\
  &=& 
  \left\langle
    \left( 
      \sum_{a\in\alpha} n_a W(|\boldsymbol{r}_a-\boldsymbol{r}_\alpha|;R_i)
    \right)^2
    \left(
      \sum_{c\in\beta} n_c W(|\boldsymbol{r}_c-\boldsymbol{r}_\beta|;R_j)
    \right)^2
  \right\rangle
  \nonumber \\
  &=&
  \langle N_i \rangle
  \langle N_j \rangle
  \left( \overline{W} \right)^2 +
  \langle N_i \rangle
  \langle N_j \rangle^2
  \left(
    1 + \overline{\xi}_{R_j}^{(0,0)}
  \right) +
  \langle N_i \rangle^2
  \langle N_j \rangle 
  \left(
    1 + \overline{\xi}_{R_i}^{(0,0)}
  \right)
  \nonumber \\
  && +
  \langle N_i \rangle^2
  \langle N_j \rangle^2
  \left(
    1 +
    \overline{\xi}_{R_i}^{(0,0)} + 
    \overline{\xi}_{R_j}^{(0,0)} +
    \overline{\xi}_{R_i}^{(0,0)}
    \overline{\xi}_{R_j}^{(0,0)}
  \right) +
  \langle N_i \rangle
  \langle N_j \rangle
  \left( \overline{W} \right)^2
  \overline{\xi}_{\alpha\beta}^{(1,1)}
  \nonumber \\
  && +
  \langle N_i \rangle
  \langle N_j \rangle^2 
  \left(
    2 \overline{\xi}_{\alpha\beta}^{(1,0)} +
    \zeta_{\alpha\beta\beta}^{(1,0,0)}
  \right) +
  \langle N_i \rangle^2
  \langle N_j \rangle
  \overline{W}
  \left(
    2\overline{\xi}_{\alpha\beta}^{(0,1)} +
    \zeta_{\alpha\alpha\beta}^{(0,0,1)}
  \right)
  \nonumber \\
  && +
  \langle N_i \rangle^2
  \langle N_j \rangle^2 
  \left[
    4 \overline{\xi}_{\alpha\beta}^{(0,0)} +
    2 \left(
      \zeta_{\alpha\alpha\beta}^{(0,0,0)} +
      \zeta_{\alpha\beta\beta}^{(0,0,0)}
    \right) +
    2 \left( \overline{\xi}_{\alpha\beta}^{(0,0)} \right)^2 +
    \overline{\eta}_{\alpha\alpha\beta\beta}^{(0,0,0,0)}
  \right].
\label{eq:B10}
\end{eqnarray}

\begin{figure}[h]
  \begin{center}
    \FigureFile(105mm,75mm){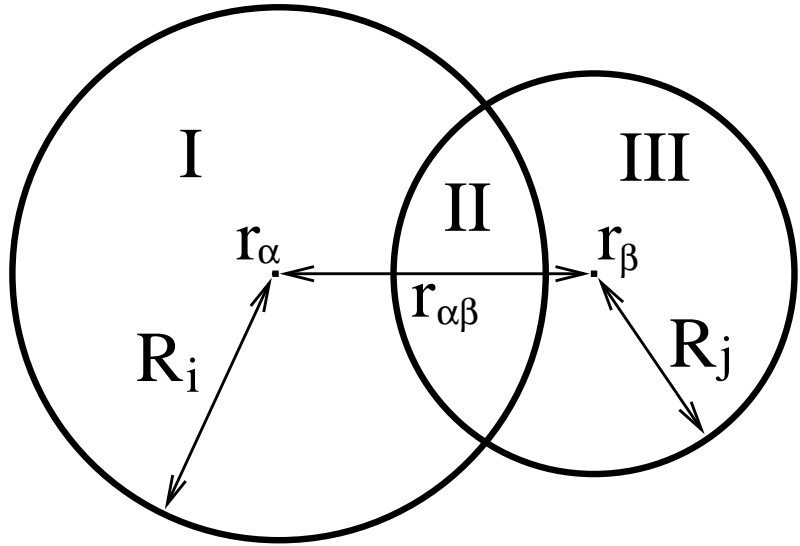}
  \end{center}
  \caption{Schematic picture of two overlapping cells whose centers
    are $\boldsymbol{r}_\alpha$ for a cell whose radius is $R_i$ and
    $\boldsymbol{r}_\beta$ for a cell whose radius is
    $R_j$. $r_{\alpha\beta}=|\boldsymbol{r}_\alpha-\boldsymbol{r}_\beta|$
    represent a separation between two centers. Region I contains cell
    $\alpha$ except overlapping region and region II is the
    overlapping region and region III contains cell $\beta$ excepting
    overlapping region. \label{fig:fig7}}
\end{figure}

On the other hand, for cell pairs that are overlapped to each other,
we calculate by following \citet{over,lap}. A configuration of cells
are shown in figure~\ref{fig:fig7}. We assume that the galaxy
distribution within overlapped cells is locally Poisson process. We
divide each cell into infinitesimal volumes as before and consider two
cases depending on whether each infinitesimal volume $\delta v_a$ is
included in a overlapped region or not. Then, the summations in
equation~(\ref{eq:B10}) is divided into
\begin{eqnarray}
  \sum_{a,b\in\alpha} &=&
  \sum_{a,b\in {\rm I}} +
  \sum_{a,b\in {\rm I\hspace{-.1em}I}} +
  2 \sum_{a\in {\rm I},b\in {\rm I\hspace{-.1em}I}}, \\
  \sum_{c,d\in\beta} &=&
  \sum_{c,d\in {\rm I\hspace{-.1em}I}} +
  \sum_{c,d\in {\rm I\hspace{-.1em}I\hspace{-.1em}I }} +
  2 \sum_{c\in {\rm I\hspace{-.1em}I},d\in {\rm I\hspace{-.1em}I\hspace{-.1em}I }}.
\end{eqnarray}
We introduce other notations to show the results
\begin{eqnarray}
  \overline{V}_{\rm I}^{(p,q)}
  &\equiv&
  \int_{\rm I} d^3r_a \,
  \{ W(|\boldsymbol{r}_a|;R_i) \}^p
  \{ W(|\boldsymbol{r}_a-\boldsymbol{r}_{\alpha\beta}|;R_j) \}^q, 
  \label{eq:B13} \\
  \overline{V}_{\rm I\hspace{-.1em}I}^{(p,q)}
  &\equiv&
  \int_{\rm I\hspace{-.1em}I} d^3r_a\,
  \{ W(|\boldsymbol{r}_a|;R_i) \}^p
  \{ W(|\boldsymbol{r}_a-\boldsymbol{r}_{\alpha\beta}|;R_j) \}^q, 
  \label{eq:B14} \\
  \overline{V}_{\rm I\hspace{-.1em}I\hspace{-.1em}I}^{(p,q)}
  &\equiv&
  \int_{\rm I\hspace{-.1em}I\hspace{-.1em}I} d^3r_a\,
  \{ W(|\boldsymbol{r}_a|;R_i) \}^p
  \{ W(|\boldsymbol{r}_a-\boldsymbol{r}_{\alpha\beta}|;R_j) \}^q,
  \label{eq:B15}
\end{eqnarray}
where each subscript I, I\hspace{-.1em}I, and
I\hspace{-.1em}I\hspace{-.1em}I represents an integration over each
region I, I\hspace{-.1em}I, and I\hspace{-.1em}I\hspace{-.1em}I,
respectively, and $\boldsymbol{r}_{\alpha\beta} =
\boldsymbol{r}_\alpha-\boldsymbol{r}_\beta$. After some calculations,
we obtain
\begin{eqnarray}
  && \hspace{-1em} \left\langle {N_{i,\alpha}}^2
    {N_{j,\beta}}^2 \right\rangle_{\rm overlap}
  \nonumber \\
  &=&
  \overline{n}\overline{V}_{\rm I\hspace{-.1em}I}^{(2,2)}
  \nonumber \\
  && +
  \overline{n}^2
  \left[
    \overline{V}_{\rm I}^{(2,0)}
    \overline{V}_{\rm I\hspace{-.1em}I}^{(0,2)} +
    \overline{V}_{\rm I}^{(2,0)}
    \overline{V}_{\rm I\hspace{-.1em}I\hspace{-.1em}I }^{(0,2)} +
    \overline{V}_{\rm I\hspace{-.1em}I}^{(2,0)}
    \overline{V}_{\rm I\hspace{-.1em}I\hspace{-.1em}I }^{(0,2)} +
    2 \overline{V}_{\rm I\hspace{-.1em}I}^{(2,1)}
    \overline{V}_{\rm I\hspace{-.1em}I}^{(0,1)} 
  \right.
  \nonumber \\
  && \hspace{1em} +
  2 \overline{V}_{\rm I\hspace{-.1em}I}^{(1,0)}
  \overline{V}_{\rm I\hspace{-.1em}I}^{(1,2)} +
  \overline{V}_{\rm I\hspace{-.1em}I}^{(2,0)}
  \overline{V}_{\rm I\hspace{-.1em}I}^{(0,2)} +
  2 \left( \overline{V}_{\rm I\hspace{-.1em}I}^{(1,1)} \right)^2 +
  \left.
    2 \overline{V}_{\rm I\hspace{-.1em}I}^{(2,1)}
    \overline{V}_{\rm I\hspace{-.1em}I\hspace{-.1em}I }^{(0,1)} +
    2 \overline{V}_{\rm I}^{(1,0)}
    \overline{V}_{\rm I\hspace{-.1em}I}^{(1,2)}
  \right]
  \nonumber \\
  && +
  \overline{n}^3
  \left[
    \left( \overline{V}_{\rm I}^{(1,0)} \right)^2
    \overline{V}_{\rm I\hspace{-.1em}I}^{(0,2)} +
    \overline{V}_{\rm I}^{(2,0)} 
    \left( \overline{V}_{\rm I\hspace{-.1em}I}^{(0,1)} \right)^2 +
    \left( \overline{V}_{\rm I}^{(1,0)} \right)^2
    \overline{V}_{\rm I\hspace{-.1em}I\hspace{-.1em}I }^{(0,2)} +
    \overline{V}_{\rm I}^{(2,0)}
    \left( \overline{V}_{\rm I\hspace{-.1em}I\hspace{-.1em}I }^{(0,1)} \right)^2
  \right .
  \nonumber \\
  && \hspace{1em} +
  \left( \overline{V}_{\rm I\hspace{-.1em}I}^{(1,0)} \right)^2
  \overline{V}_{\rm I\hspace{-.1em}I\hspace{-.1em}I }^{(0,2)} +
  \overline{V}_{\rm I\hspace{-.1em}I}^{(2,0)}
  \left( \overline{V}_{\rm I\hspace{-.1em}I\hspace{-.1em}I }^{(0,1)} \right)^2 +
  2 \overline{V}_{\rm I}^{(2,0)}
  \overline{V}_{\rm I\hspace{-.1em}I}^{(0,1)}
  \overline{V}_{\rm I\hspace{-.1em}I\hspace{-.1em}I }^{(0,1)} +
  2 \overline{V}_{\rm I}^{(1,0)}
  \overline{V}_{\rm I\hspace{-.1em}I}^{(1,0)}
  \overline{V}_{\rm I\hspace{-.1em}I\hspace{-.1em}I }^{(0,2)}
  \nonumber \\
  && \hspace{1em} +
  4 \overline{V}_{\rm I}^{(1,0)}
  \overline{V}_{\rm I\hspace{-.1em}I}^{(1,1)}
  \overline{V}_{\rm I\hspace{-.1em}I\hspace{-.1em}I }^{(0,1)} +
  2 \overline{V}_{\rm I\hspace{-.1em}I}^{(2,0)}
  \left(\overline{V}_{\rm I\hspace{-.1em}I}^{(0,1)}\right)^2 +
  2 \overline{V}_{\rm I\hspace{-.1em}I}^{(1,1)}
  \overline{V}_{\rm I\hspace{-.1em}I}^{(1,0)}
  \overline{V}_{\rm I\hspace{-.1em}I}^{(0,1)} +
  2 \left(\overline{V}_{\rm I\hspace{-.1em}I}^{(1,0)}\right)^2
  \overline{V}_{\rm I\hspace{-.1em}I}^{(0,2)}
  \nonumber \\
  && \hspace{1em} +
  \left.
    2 \overline{V}_{\rm I\hspace{-.1em}I}^{(2,0)}
    \overline{V}_{\rm I\hspace{-.1em}I}^{(0,1)}
    \overline{V}_{\rm I\hspace{-.1em}I\hspace{-.1em}I }^{(0,1)} +
    4 \overline{V}_{\rm I\hspace{-.1em}I}^{(1,1)}
    \overline{V}_{\rm I\hspace{-.1em}I}^{(1,0)}
    \overline{V}_{\rm I\hspace{-.1em}I\hspace{-.1em}I }^{(0,1)} +
    2 \overline{V}_{\rm I}^{(1,0)}
    \overline{V}_{\rm I\hspace{-.1em}I}^{(1,0)}
    \overline{V}_{\rm I\hspace{-.1em}I}^{(0,2)}
  \right.
  \nonumber \\
  && \hspace{1em} +
  \left.
    4 \overline{V}_{\rm I}^{(1,0)}
    \overline{V}_{\rm I\hspace{-.1em}I}^{(1,1)}
    \overline{V}_{\rm I\hspace{-.1em}I}^{(0,0)}
  \right]
  \nonumber \\
  && +
  \overline{n}^4
  \left[
    \left(
      \overline{V}_{\rm I}^{(1,0)}
      \overline{V}_{\rm I\hspace{-.1em}I}^{(0,2)}
    \right)^2 +
    \left(
      \overline{V}_{\rm I}^{(1,0)}
      \overline{V}_{\rm I\hspace{-.1em}I\hspace{-.1em}I }^{(0,1)}
    \right)^2 +
    \left(
      \overline{V}_{\rm I\hspace{-.1em}I}^{(1,0)}
      \overline{V}_{\rm I\hspace{-.1em}I\hspace{-.1em}I }^{(0,1)}
    \right)^2 +
    2\left( \overline{V}_{\rm I}^{(1,0)} \right)^2
    \overline{V}_{\rm I\hspace{-.1em}I}^{(0,1)}
    \overline{V}_{\rm I\hspace{-.1em}I\hspace{-.1em}I }^{(0,1)}
  \right.
  \nonumber \\
  && \hspace{1em} +
  2\overline{V}_{\rm I}^{(1,0)}
  \overline{V}_{\rm I\hspace{-.1em}I}^{(1,0)}
  \left(
    \overline{V}_{\rm I\hspace{-.1em}I\hspace{-.1em}I }^{(0,1)}
  \right)^2 +
  4\overline{V}_{\rm I}^{(1,0)}
  \overline{V}_{\rm I\hspace{-.1em}I}^{(1,0)}
  \overline{V}_{\rm I\hspace{-.1em}I}^{(0,1)}
  \overline{V}_{\rm I\hspace{-.1em}I\hspace{-.1em}I }^{(0,1)} +
  \left(
    \overline{V}_{\rm I\hspace{-.1em}I}^{(1,0)}
    \overline{V}_{\rm I\hspace{-.1em}I}^{(0,1)}
  \right)^2
  \nonumber \\
  && \hspace{1em} +
  \left.
    2 \left( \overline{V}_{\rm I\hspace{-.1em}I}^{(1,0)} \right)^2
    \overline{V}_{\rm I\hspace{-.1em}I}^{(0,1)}
    \overline{V}_{\rm I\hspace{-.1em}I\hspace{-.1em}I }^{(0,1)} +
    2\overline{V}_{\rm I}^{(1,0)}
    \overline{V}_{\rm I\hspace{-.1em}I}^{(1,0)}
    \left( \overline{V}_{\rm I\hspace{-.1em}I}^{(0,1)} \right)^2 
  \right].
\label{eq:B16}
\end{eqnarray}

It is straightforward to evaluate $\overline{V}$ terms that are
integrated over region ${\rm I}\hspace{-.1em}{\rm I}$ in the above
equation. For example, by using the Fourier transformation, we have
\begin{eqnarray}
  \overline{V}_{{\rm I}\hspace{-.1em}{\rm I}}^{(p,q)}
  &=&
  \int d^3r_a \, 
  \{ W(|\boldsymbol{r}_a|;R_i) \}^p
  \{ W(|\boldsymbol{r}_a-\boldsymbol{r}_{\alpha\beta}|;R_j) \}^q
  \nonumber \\
  &=&
  \int \frac{d^3k}{(2\pi)^3} \,
  \tilde{W}^{(p)}(kR_i)
  \tilde{W}^{(q)}(kR_j)
  \frac{\sin kr_{\alpha\beta}}{kr_{\alpha\beta}},
 \end{eqnarray}
in an overlapped region, where
\begin{equation}
  \tilde{W}^{(p)}(kR)
  \equiv
  \int d^3r \,
  \left\{ W(|\boldsymbol{r}|;R) \right\}^p
  e^{-i\boldsymbol{k}\cdot\boldsymbol{r}}.
\end{equation}
On the other hand, for $\overline{V}$ terms that are integrated over
non-overlapping region, it is not a good way to evaluate
equation~(\ref{eq:B13}) and (\ref{eq:B15}) directly, because it seems
we need numerical calculations in this expression. However, we can
analytically subtract a contribution of the overlapping region from an
integration over a whole cell. First, note that an integration over a
non-overlapped region has nothing to do with a shape of an overlapping
cell. Therefore, the contribution from the overlapped region can be
calculated by replacing the overlapping cell by the top-hat smoothing
function having same center and radius as the overlapping cell. Then,
we obtain
\begin{eqnarray}
  \overline{V}_{\rm I}^{(p,q)}
  &=&
  \int d^3r_a\, \{W(|\boldsymbol{r}_a|;R_i)\}^p -
  \int d^3r_a\, \{W(|\boldsymbol{r}_a|;R_i)\}^p
  \Theta(R_j-|\boldsymbol{r}_a-\boldsymbol{r}_{\alpha\beta}|)
  \nonumber \\
  &=&
  v_i \overline{W^{p-1}} - 
  \int \frac{d^3k}{(2\pi)^3} \,
  \tilde{W}^{(p)}(kR_i) 
  \left\{v_j \tilde{U}_{\rm TH}(kR_j) \right\}
  \frac{\sin kr_{\alpha\beta}}{kr_{\alpha\beta}}, 
  \\
  \overline{V}_{{\rm I}\hspace{-.1em}{\rm I}\hspace{-.1em}{\rm I}}^{(p,q)}
  &=&
  \int d^3r_a \, \{ W(|\boldsymbol{r}_a|;R_j) \}^q -
  \int d^3r_a \, \{ W(|\boldsymbol{r}_a|;R_j) \}^q
  \Theta(R_i-|\boldsymbol{r}_a-\boldsymbol{r}_{\alpha\beta}|)
  \label{A15}
  \nonumber \\
  &=&
  v_j \overline{W^{q-1}} -
  \int \frac{d^3k}{(2\pi)^3}\,
  \left\{ v_i \tilde{U}_{\rm TH}(kR_i) \right\}
  \tilde{W}^{(q)}(kR_j)
  \frac{\sin kr_{\alpha\beta}}{kr_{\alpha\beta}},
\label{eq:B20}
\end{eqnarray}
where $v_i=4\pi R_i^3/3$, $v_j=4\pi R_j^3/3$, and
$\tilde{U}_{\rm TH}(x)=3(\sin x-x\cos x)/x^3$ is the Fourier
transformation of the top-hat function.

In the case of the $m$-weight Epanechnikov kernel, for example, we can
use equation~(\ref{eq:110}) for $p=1$ or $q=1$. For $p=2$ or $q=2$, a
square of the $m$-weight Epanechnikov kernel is related to the
$2m$-weight Epanechnikov kernel by a following relation:
\begin{equation}
  \left\{W_m(|\boldsymbol{r}|;R)\right\}^2
  = \frac{1}{3} \left\{\frac{(2m+3)!!}{m!}\right\}^2 \frac{(2m)!}{(4m+3)!!}
  W_{2m}(|\boldsymbol{r}|;R).
\end{equation}
Therefore, we can use
\begin{equation}
  \tilde{W}_m^{(2)}(kR)
  = \frac{1}{3} \left\{\frac{(2m+3)!!}{m!}\right\}^2 \frac{(2m)!}{(4m+3)!!}
  \tilde{W}_{2m}(kR).
\end{equation}

By using equations~(\ref{eq:B2}), (\ref{eq:B3}), (\ref{eq:B8}) -
(\ref{eq:B10}), (\ref{eq:B16}) - (\ref{eq:B20}) and by approximating
$1\gg\xi\gg\zeta\sim\xi^2\gg\eta\sim\xi^3$, we finally obtain
\begin{eqnarray}
  C_{ii}
  &=&
  \frac{1}{N_{{\rm cell},i}}
  \left[
    \frac{\overline{W^3}}{\langle N_i \rangle^3} +
    4 \frac{\overline{W^2}}{\langle N_i \rangle^2} +
    2 \frac{\left(\overline{W}\right)^2}{\langle N_i \rangle^2} +
    4 \frac{\overline{W}}{\langle N_i \rangle}
  \right]
  \nonumber \\
  && +
  \frac{1}{V^2} 
  \left(
    1 - \frac{1}{N_{{\rm cell},i}}
  \right)
  \int_{r_{\alpha\beta}\le 2R_i} d^2r_\alpha d^3r_\beta \,
  \frac{1}{\langle N_i \rangle^4} 
  \left\langle {N_{i,\alpha}}^2{N_{i,\beta}}^2 \right\rangle_{\rm overlap}
  \nonumber \\
  && +
  \frac{1}{V^2} 
  \left(
    1 - \frac{1}{N_{{\rm cell},i}}
  \right)
  \int_{r_{\alpha\beta}>2R_i}d^3r_\alpha d^3r_\beta\,
  \left[
    \frac{\left(\overline{W}\right)^2}{\langle N_i\rangle^2}
    \overline{\xi}_{\alpha\beta}^{(1,1)} +
    4\frac{\overline{W}}{\langle N_i\rangle}
    \overline{\xi}_{\alpha\beta}^{(1,0)} +
    4\overline{\xi}_{\alpha\beta}^{(0,0)}
  \right],
  \nonumber \\
  \label{eq:B21} \\
  C_{ij}
  &=&
  \frac{1}{V^2}
  \int_{r_{\alpha\beta}\le R_i+R_j} d^3r_\alpha d^3r_\beta \,
  \frac{1}{\langle N_i \rangle^2 \langle N_j \rangle^2}
  \left\langle
    {N_{i,\alpha}}^2 {N_{j,\beta}}^2
  \right\rangle_{\rm overlap}
  \nonumber \\
  && +
  \frac{1}{V^2}
  \int_{r_{\alpha\beta}>R_i+R_j}d^3r_\alpha d^3r_\beta\,
  \left[
    \frac{\left(\overline{W}\right)^2}{\langle N_i\rangle\langle N_j\rangle}
    \overline{\xi}_{\alpha\beta}^{(1,1)} +
    2\left(
      \frac{\overline{W}}{\langle N_i\rangle}
      \overline{\xi}_{\alpha\beta}^{(1,0)} +
      \frac{\overline{W}}{\langle N_j\rangle}
      \overline{\xi}_{\alpha\beta}^{(0,1)}
    \right) +
    4\overline{\xi}_{\alpha\beta}^{(0,0)}
  \right],
  \nonumber \\
  \label{eq:B22}
\end{eqnarray}
where
\begin{eqnarray}
  &&  \hspace{-1em} \frac{1}{V^2}
  \int_{r_{\alpha\beta}\le R_i+R_j} d^3r_\alpha d^3r_\beta \,
  \frac{1}{\langle N_i \rangle^2 \langle N_j \rangle^2}
  \left\langle
    {N_{i,\alpha}}^2 {N_{j,\beta}}^2
  \right\rangle_{\rm overlap}
  \nonumber \\
  &=&
  \frac{1}{\overline{n}^3{v_i}^2{v_j}^2}
  \left( \frac{R_i+R_j}{L} \right)^3
  \int \frac{d^3k}{(2\pi)^3} \,
  \tilde{W}^{(2)}(kR_i) \tilde{W}^{(2)}(kR_j) \tilde{U}_{\rm TH}
  \left( k\left( R_i+R_j\right)
  \right)
  \nonumber \\
  && +
  \frac{2}{\overline{n}^2{v_i}^2v_j}
  \left( \frac{R_i+R_j}{L} \right)^3
  \int \frac{d^3k}{(2\pi)^3} \,
  \tilde{W}^{(2)}(kR_i) \tilde{W}(kR_j) \tilde{U}_{\rm TH}
  \left( k \left( R_i+R_j \right) \right)
  \nonumber \\
  && +
  \frac{2}{\overline{n}^2v_i{v_j}^2}
  \left( \frac{R_i+R_j}{L} \right)^3
  \int \frac{d^3k}{(2\pi)^3} \,
  \tilde{W}(kR_i) \tilde{W}^{(2)}(kR_j) \tilde{U}_{\rm TH}
  \left( k \left( R_i+R_j \right) \right)
  \nonumber \\
  && +
  \frac{4}{\overline{n}v_iv_j}
  \left( \frac{R_i+R_j}{L} \right)^3
  \int \frac{d^3k}{(2\pi)^3} \,
  \tilde{W}(kR_i) \tilde{W}(kR_j) \tilde{U}_{\rm TH}
  \left( k \left( R_i + R_j \right) \right)
  \nonumber \\
  && +
  \frac{2}{\overline{n}^2{v_i}^2{v_j}^2}
  \frac{1}{V} \int^{R_i+R_j}_0 4 \pi r^2dr \,
  \left(
    \int \frac{d^3k}{(2\pi)^3} \,
    \tilde{W}(kR_i) \tilde{W}(kR_j) \frac{\sin kr}{kr}
  \right)^2
  \nonumber \\
  && +
  \frac{1}{\overline{n}{v_i}^2{v_j}^2}
  \frac{1}{V} \int^{R_i+R_j}_04\pi r^2dr
  \left[
    \left(
      \int \frac{d^3k_1}{(2\pi)^3} \,
      \tilde{W}(k_1R_i)
      \left\{ v_j \tilde{U}_{\rm TH}(k_1R_j) \right\}
      \frac{\sin k_1r}{k_1r}
    \right)^2
  \right.
  \nonumber \\
  && \hspace{13em} \times
  \left.
    \int \frac{d^3k_2}{(2\pi)^3} \,
    \left\{ v_i \tilde{U}_{\rm TH}(k_2R_i) \right\}
    \tilde{W}^{(2)}(k_2R_j) \frac{\sin k_2r}{k_2r}
  \right.
  \nonumber \\
  && \hspace{12em} +
  \left.
    \int \frac{d^3k_1}{(2\pi)^3} \,
    \tilde{W}^{(2)}(k_1R_i) 
    \left\{ v_j \tilde{U}_{\rm TH}(k_1R_j) \right\}
    \frac{\sin k_1r}{k_1r}
  \right.
  \nonumber \\
  && \hspace{13em} \times
  \left.
    \left(
      \int \frac{d^3k_2}{(2\pi)^3} \,
      \left\{ v_i \tilde{U}_{\rm TH}(k_2R_i) \right\}
      \tilde{W}(k_2R_j) \frac{\sin k_2r}{k_2r}
    \right)^2
  \right]
  \nonumber \\
  && -
  \frac{2}{\overline{n}{v_i}^2{v_j}^2}
  \frac{1}{V}
  \int^{R_i+R_j}_04\pi r^2dr \,
  \left[
    \int \frac{d^3k_1}{(2\pi)^3} \,
    \tilde{W}(k_1R_i)
    \left\{ v_j \tilde{U}_{\rm TH}(k_1R_j) \right\}
    \frac{\sin k_1r}{k_1r}
  \right.
  \nonumber \\
  && \hspace{1em} \times
  \left.
    \int \frac{d^3k_2}{(2\pi)^3} \,
    \tilde{W}(k_2R_i) \tilde{W}(k_2R_j)
    \frac{\sin k_2r}{k_2r}
    \int \frac{d^3k_3}{(2\pi)^3} \,
    \left\{ v_i \tilde{U}_{\rm TH}(k_3R_i) \right\}
    \tilde{W}(k_3R_j) \frac{\sin k_3r}{k_3r}
  \right].
  \nonumber \\
  \label{eq:B23}
\end{eqnarray}

\section{Convenient Forms of Equations~(\ref{eq:same}) and
  (\ref{eq:diff}) for Numerical Calculation}

To calculate equations~(\ref{eq:same}) and (\ref{eq:diff})
numerically, it is convenient to go forward more analytically. By
using the inverse Fourier transformation
\begin{equation}
  \left\{ W(|\boldsymbol{r}|;R) \right\}^n =
  \int \frac{d^3k}{(2\pi)^3} \,
  \tilde{W}^{(n)}(kR)
  e^{i\boldsymbol{k}\cdot\boldsymbol{r}},
\end{equation}
equation~(\ref{eq:same}) becomes
\begin{equation}
  \overline{\xi}^{(p,r)}_R =
  \int \frac{d^3k}{(2\pi)^3} \,
  \frac{\tilde{W}^{(p+1)}(kR)}{v\overline{W^p}}
  \frac{\tilde{W}^{(q+1)}(kR)}{v\overline{W^q}}
  P(k)
  \label{eq:xibarsame},
\end{equation}
where $v=4\pi R^3/3$. Similarly, equation~(\ref{eq:diff}) becomes
\begin{equation}
  \overline{\xi}^{(p,r)}_{\alpha\beta} =
  \int \frac{d^3k}{(2\pi)^3} \,
  \frac{\tilde{W}^{(p+1)}(kR_i)}{v_i\overline{W^p}}
  \frac{\tilde{W}^{(q+1)}(kR_j)}{v_j\overline{W^q}}
  P(k) e^{-i\boldsymbol{k}\cdot(\boldsymbol{r}_\alpha-\boldsymbol{r}_\beta)}
\label{eq:xibarF}.
\end{equation}
Then, the integration over non-overlapped region appeared in equation
(\ref{eq:B22}) becomes
\begin{eqnarray}
  && \hspace{-1em} \frac{1}{V^2}
  \int_{r_{\alpha\beta}>R_i+R_j}d^3r_\alpha d^3r_\beta \,
  \overline{\xi}_{\alpha\beta}^{(p,q)}
  \nonumber\\
  &=&
  \int \frac{d^3k}{(2\pi)^3} \,
  \frac{\tilde{W}^{(p+1)}(kR_i)}{v_i\overline{W^p}}
  \frac{\tilde{W}^{(q+1)}(kR_j)}{v_j\overline{W^q}}
  P(k)
  \left[
    \tilde{U}_{\rm TH}(kL) -
    \left( \frac{R_i+R_j}{L} \right)^3
    \tilde{U}_{\rm TH}\left(k(R_i+R_j)\right)
  \right],
  \nonumber \\
  \label{eq:xibardiff}
\end{eqnarray}
because we assume that a survey volume has a spherical shape.

\clearpage

\clearpage

\clearpage

\clearpage

\clearpage

\clearpage

\clearpage

\clearpage

\clearpage

\end{document}